\documentstyle[12pt]{article}
\input{epsf}
\newcommand{\be}{\begin{eqnarray}}
\newcommand{\ee}{\end{eqnarray}}
\newcommand{\ba}{\begin{array}}
\newcommand{\ea}{\end{array}}

\newcommand{\cM}{{\cal M}}

\newcommand{\partialslash}{\partial\hspace{-.5em}/\hspace{.15em}}
\newcommand{\pz}{p\!\cdot\! z}
\newcommand{\rz}{r\!\cdot\! z}

\newcommand{\nablaboth}{\stackrel{\leftrightarrow}{\nabla}}
\begin{document}
%
%
\rightline{RUB-TPII-1/99}
\rightline{hep-ph/9902451}
\vspace{.3cm}
\begin{center}
\begin{large}
{\bf Skewed and double distributions in pion and nucleon}
\end{large}
\\[1cm]
{\bf M.V.\ Polyakov$^{\rm 1, 2, a}$ and C.\ Weiss$^{\rm 2, b}$}
\\[0.2cm]
$^{\rm 1}${\em Theory Division of Petersburg Nuclear Physics Institute \\
188350 Gatchina, Leningrad District, Russian Federation}\\
\vspace{0.2cm}
$^{\rm 2}${\em Institut f\"ur Theoretische Physik II \\
Ruhr--Universit\"at Bochum \\
D--44780 Bochum, Germany}
\end{center}
\vspace{1cm}
\begin{abstract}
\noindent
We study the non-forward matrix elements of twist--2 QCD
light--ray operators and their representations in terms of skewed 
and double distributions, considering the pion as well as the 
nucleon. We point out the importance of explicitly including all 
twist--2 structures in the double distribution representation, which
naturally leads to a ``two--component'' structure of the skewed 
distribution, with different contributions in the regions 
$|X| > \xi / 2$ and $|X| < \xi / 2$. We compute
the skewed and double quark distributions in the pion at a low 
normalization point in the effective chiral theory 
based on the instanton vacuum. Also, we derive the crossing
relations expressing the skewed quark distribution in the pion 
through the distribution amplitude for two--pion production. 
Measurement of the latter in two--pion production in 
$\gamma^\ast \gamma$ and $\gamma^\ast N$ reactions could provide 
direct information about the skewed as well as the usual 
quark/antiquark--distribution in the pion.
\end{abstract}
\vspace{1.5cm} PACS: 12.38.Lg, 13.60.Fz, 13.60.Le
\\
Keywords: \parbox[t]{13cm}{skewed parton distributions,
meson distribution amplitudes, chiral symmetry,
non-perturbative methods in QCD}
\vfill
\rule{5cm}{.15mm}
\\
\noindent
{\footnotesize $^{\rm a}$ E-mail: maximp@tp2.ruhr-uni-bochum.de} \\
{\footnotesize $^{\rm b}$ E-mail: weiss@tp2.ruhr-uni-bochum.de}
%
%
%
%
\newpage 
\tableofcontents
%
%
%
\newpage
\section{Introduction}
In so--called light--cone dominated hard scattering processes the
non-perturbative information entering the scattering amplitude
is contained in matrix elements of certain QCD light--ray operators
between hadronic states. A well--known example is inclusive
deep--inelastic scattering, where in the asymptotic regime
the cross section is determined by forward (diagonal) matrix
elements of twist--2 light--ray operators in the target state,
which have an interpretation as parton distributions. More recently,
factorization has been proven for a large class of exclusive
processes, namely deeply--virtual Compton scattering (DVCS) 
and hard meson production \cite{CFS97,Ji1,Ji2,Rad1,Rad2,JO97,CF98,GV98}. 
The amplitudes for these processes involve
non-forward (more generally, non-diagonal) matrix elements
of light--ray operators between incoming and outgoing hadron 
states, which can be represented by generalized parton 
distributions. Such matrix elements had earlier been
introduced in the description of $Z^0$ photoproduction at small $x$
\cite{BartelsLoewe82} and in the context of the non-local
light--cone expansion \cite{Geyer85}.
\par
Due to the presence of a non-zero momentum transfer, non-forward
matrix elements of light--ray operators possess a much richer structure
than the forward ones defining the familiar parton distributions.
In the matrix element of a generic light--ray operator,
$\langle p - r/2 | \varphi(-z/2) \ldots \varphi(z/2) | p + r/2 \rangle$,
with $z^2 = 0$, both the momentum transfer, $r$, and the average of initial 
and final momenta, $p$, in general have non-zero longitudinal (``plus'')
component with respect to the light--cone direction defined by $z$. 
In a partonic language, one may express the momentum of the 
``active'' parton in terms of any linear combination of $\pz$ and $\rz$.
Two approaches have been proposed. One can analyze the
matrix element assuming proportionality $\rz = \xi \,\pz$, where
the value of $\xi$ is determined by the kinematics of the
scattering process ({\it e.g.}\ in DVCS it is related to the Bjorken
variable). This leads to the so-called skewed distributions\footnote{The 
term ``skewed distribution'' has been recommended as a common 
name for the ``off-forward'' distributions introduced by 
Ji \cite{Ji1,Ji2} and the ``non-forward'' distributions of
Radyushkin \cite{Rad1,Rad2}, which differ in the 
definition of the parton momenta, see Ref.\cite{Rad2} for a detailed
discussion. It encompasses
also the ``non-diagonal'' distributions parametrizing matrix elements
between hadron states of different quantum numbers, 
as have been introduced {\it e.g.}\ to describe DVCS with 
$N$--$\Delta$ transitions \cite{FPS98}.},
families of generalized parton distributions depending explicitly on the
``skewedness'' parameter, $\xi$ \cite{Ji1,Ji2,Rad1,Rad2}. In another 
approach, proposed by Radyushkin \cite{Rad1,Rad2}, one writes a spectral 
representation for the matrix element of the light--ray operator 
as an independent function of $\pz$ and $\rz$ in terms of a so-called 
double distribution. The skewed distribution for a given value of $\xi$
is then obtained as a particular one--dimensional reduction 
of this two--variable distribution. The advantage of this approach is 
that it allows one to make statements about the dependence of the skewed 
distribution on the skewedness parameter, $\xi$.
\par
The general structure of skewed and double 
distributions --- their symmetries, limiting cases, possible 
singularities, {\it etc.} --- is a problem of great theoretical and 
practical importance. This problem has two aspects. The distributions 
depend, of course, on the behavior of the matrix elements 
of the light--ray operators as functions of $\pz , \rz$. This is a 
dynamical question, which one
can address from the point of view of general invariance principles, 
or by calculations using some dynamical model. However, the properties 
of the distributions are also determined by the particular way in which 
one writes the spectral representation for the matrix element.
This concerns such things as {\it e.g.}\ the number of independent 
``twist--2 structures'' one includes in the double distribution 
representation of the matrix element. A clear understanding of
both aspects of this problem is necessary for building realistic 
models for skewed distributions.
\par
In this paper we investigate the structure of hadronic matrix elements 
of twist--2 operators at a low normalization point, using
general principles (symmetries, crossing {\it etc}.) as well as specific 
dynamical models, and consider 
the implications for spectral representations in terms of skewed and 
double distributions. We show that the standard definition of the double 
distribution representation of Refs.\cite{Rad1,Rad2} is not always 
compatible with the basic structure of the matrix elements
(to the very least, it implies severe singularities of the
double distribution), and propose a complete representation which 
explicitly takes into account all twist--2 structures. 
The additional terms give rise to contributions to the skewed 
distribution which are non-zero only in the region 
$-\xi / 2 < X < \xi / 2$, and thus naturally lead to a 
``two--component'' form of the skewed distribution, {\it i.e.}, 
to essentially different functions in the two regions 
$|X| < \xi/2$ and $|X| > \xi/2$. Such behavior was first observed 
in a model calculation of the flavor--singlet skewed distribution 
in the large--$N_c$ limit in Ref.\cite{PPPBGW97}.
\par
We find it useful to consider in addition to the nucleon matrix 
elements of twist--2 light--ray operators also the matrix elements 
between pion states. While hardly the target of choice for actual 
DVCS experiments, the pion is interesting from a theoretical 
point of view, for various reasons. First, it allows
to avoid complications due to spin, and also its mass can be
neglected. Second, the interactions of the pion with
external fields are described completely by the chiral Lagrangian, 
which makes it possible to derive certain sum rules for the skewed 
distributions at a low normalization point from first principles. 
Finally, both the skewed and the double distribution in the pion at a 
low normalization point can be estimated in the large--$N_c$ limit 
in the effective low--energy theory derived from the instanton vacuum of 
QCD \cite{DP86}. This is a fully field--theoretic description of the 
pion, which respects general properties such as crossing symmetry
{\it etc.}, and incorporates the consequences of the dynamical 
breaking of chiral symmetry. The same approach has been shown to 
give a realistic description of the quark/antiquark distributions 
in the nucleon (both usual \cite{DPPPW96} and skewed \cite{PPPBGW97})
as well as the pion distribution amplitude \cite{PP97,PPRGW98}.
In fact, the results of our calculation of the skewed and double 
distributions in the pion fully support our general conclusions
concerning the need to modify the double distribution 
representation of Refs.\cite{Rad1,Rad2} and the ``two--component''
structure of the skewed distribution.\footnote{In the 
case of the nucleon the calculation of non-forward 
matrix elements is complicated by the parametric restrictions
imposed on the different components of the nucleons' momenta
by the $1/N_c$--expansion. While it is possible to compute
within the standard $1/N_c$--expansion the
skewed distributions in the nucleon in the parametric range 
$X , \xi \sim 1/N_c$ \cite{PPPBGW97}, it is difficult 
to get the double distribution in the nucleon in this approach; 
see Section \ref{sec_model}.}
\par
Another reason for our interest in the pion is the fact that
the process related to DVCS off the pion by crossing, namely 
production of two pions in $\gamma^\ast \gamma$ collisions, can be measured 
at low invariant masses \cite{Ter}. This process is described by a 
two--pion distribution amplitude, which, by crossing, is related
to the skewed parton distribution in the pion \cite{PW98,MVP98}. 
In this paper we derive the explicit relation between the two functions,
using dispersion relations to connect the regions of spacelike and 
timelike momentum transfers. In particular, this relation allows us
to connect moments of the usual quark/antiquark distribution 
in the pion to characteristics of distribution amplitudes of two--pion 
resonances \cite{MVP98}. Given the contributions that measurements 
of $\gamma^\ast \gamma \rightarrow \pi^0$ \cite{CLEO98} have made
to our knowledge of the single--pion distribution amplitude,
study of the process $\gamma^\ast \gamma \rightarrow 2\pi$
could well be one of the cleanest ways to get information about the
quark distributions in the pion --- skewed as well as usual.
\par
The scale dependence of skewed and double distributions is described 
by generalized evolution equations, which combine features of both the 
DGLAP evolution for usual parton distributions and the 
Efremov--Radyushkin--Brodsky--Lepage
evolution \cite{ERBL} for meson distribution amplitudes. This problem 
has extensively been treated in the literature, see {\it e.g.}\ 
Refs.\cite{Ji2,Rad1,Rad2,BartelsLoewe82,Geyer85,Freund98,Belitsky98,Bl97}.
We shall not be concerned with this aspect here, but rather focus on the
structure of the distributions at a low normalization point, how they
are constrained by general principles (symmetries, crossing, {\it etc.})
and how they can be estimated in dynamical models taking into account
non-perturbative effects such as the dynamical breaking of 
chiral symmetry, {\it etc.}
\par
We shall proceed as follows. In Section \ref{sec_general} 
we discuss the properties of non-forward hadronic matrix elements of 
twist--2 operators and their spectral representation from a general point 
of view. In Subsection \ref{subsec_trouble}, using the pion as the 
simplest example, we show the importance 
of explicitly including all twist--2 structures in the double distribution 
representation, and discuss the implications for skewed distributions.
The investigation is extended to nucleon matrix elements in 
Subsection \ref{subsec_nucleon}, with analogous conclusions.
In Section \ref{sec_model} we perform a model calculation
of the non-forward pion matrix elements and the corresponding skewed
and double distributions at a low normalization point 
($\mu \sim 600\, {\rm MeV}$), using the effective low--energy 
theory based on the instanton vacuum. The results serve as an 
illustration for the general discussion in Section \ref{sec_general}.
In Section \ref{sec_crossing} we discuss the relation of the 
skewed distribution in the pion to the two--pion distribution
amplitude. The crossing relation is derived in explicit form 
using moments. We use the crossing formula, together with the dispersion
relation for the invariant--mass dependence of the two--pion distribution
amplitude, to relate moments of the pion parton distribution 
to parameters of the distribution amplitudes of two--pion resonance
wave functions. Our conclusions are summarized in 
Section \ref{sec_conclusions}.
\par
Appendix \ref{app_em} gives a derivation of the generalized momentum 
sum rule for the skewed distributions in the pion.
In Appendix \ref{app_resonance} we consider ``resonance exchange''
contributions to the non-forward matrix elements in the pion. A
general expression describing the contribution of the exchange of 
$t$--channel resonances of arbitrary spin is given. 
The results provide a simple dynamical explanation for the
general properties of skewed and double distributions 
discussed in Section \ref{sec_general}.
\section{Nonforward matrix elements and generalized parton distributions}
\label{sec_general}
\subsection{Skewed vs.\ double distributions: the pion}
\label{subsec_pion}
To begin, we would like to discuss some general properties
of non-forward hadronic matrix elements of QCD light--ray 
operators and their representation in terms of skewed and double 
distributions. We start with the simplest case, the pion, 
which already exhibits all features of interest to us here,
and generalize to the nucleon in Subsection \ref{subsec_nucleon}.
\par
Let us consider the non-forward matrix elements of twist--2
light--ray operators (normalized at some scale, $\mu$) 
between one--pion states. Due to isospin invariance,
the matrix elements of the flavor--singlet and non-singlet quark operators
are of the form\footnote{From Eq.(\ref{me_def}) the matrix elements 
in charge eigenstates are obtained in the usual way: 
$|\pi^0 \rangle \equiv |\pi^3 \rangle, \;
|\pi^\pm \rangle \equiv (|\pi^1 \rangle \pm i |\pi^2 \rangle ) / \sqrt{2}$. 
Note that the neutral pion has no non-singlet matrix element due to
$C$--invariance.}
\be
\lefteqn{ \langle \pi^a (p - r/2 ) | \; \bar\psi (-z/2 ) 
\left\{ \begin{array}{c} 1 \\[2ex] \tau^c \end{array} \right\}
\hat{z} \, [-z/2, z/2] \, \psi (z/2)
\; | \pi^b (p + r/2 ) \rangle } \hspace{3cm}
&& \nonumber \\
&=& \left\{ \begin{array}{r}
2 \delta^{ab} \; \cM^{I=0} (\pz , \,\, \rz ; \,\, t)
\\[2ex]
2 i \varepsilon^{abc} \cM^{I=1} (\pz , \,\, \rz ; \,\, t) .
\end{array} \;\;\;
\right.
\label{me_def}
\ee
Throughout the following the isoscalar and isovector parts of matrix
elements in the pion will be understood to be defined as in 
Eq.(\ref{me_def}); the isospin decomposition will often not be 
explicitly written. Here $1$ and $\tau^3$ are flavor matrices; 
we consider the $SU(2)$ flavor group. Furthermore, 
$\psi$ is the quark field, $z_\mu$ a 
light--like distance ($z^2 = 0$), and 
$\hat{z} \equiv \gamma^\mu z_\mu$. Finally, $[-z/2, z/2]$ denotes 
the path--ordered exponential of the gauge field (phase factor) 
in the fundamental representation,
\be
[z/2, -z/2] &\equiv& \mbox{P}\, \exp \left[
\int_{1/2}^{-1/2} dt\; z^\mu A_\mu (t z) ,
\right]
\label{P_exp}
\ee
which is required by gauge invariance; the path here is along the 
light--like direction, $z$. The matrix element of the corresponding 
twist--2 gluon operator is defined as
\be
\lefteqn{ z^\mu z^\nu 
\langle \pi^a (p - r/2 ) | \; F_{\mu\alpha} (-z/2 ) 
\, [-z/2, z/2] \, F^\alpha_{\;\;\nu} (z/2) \; | \pi^b (p + r/2 ) \rangle }
\hspace{3cm} 
&& \nonumber \\
&=& 2 \delta^{ab} \; \cM^G (\pz , \,\, \rz ; \,\, t) ,
\label{me_gluon}
\ee
where $F_{\mu\nu}$ denotes the gluon field, and the phase factor
is in the adjoint representation. 
\par
In Eqs.(\ref{me_def}) and (\ref{me_gluon}) the dynamical information 
is contained in scalar functions, $\cM^{I = 0, 1}$ and $\cM^G$, 
which depend on the dimensionless invariants $\pz$ and $\rz$, 
as well as of $t \equiv r^2$. From the mass shell 
conditions $(p \pm r/2 )^2 = M_\pi^2$ it follows that
\be
p\!\cdot\! r &=& 0, \hspace{2cm} p^2 \;\; = \;\; M_\pi^2 - \frac{t}{4} ,
\ee
so $t$ is the only independent dimensionful invariant.
In the physical region $t < 0$. In the following we consider the
massless limit, $M_\pi \rightarrow 0$. We note that 
$G$--parity (or, equivalently, time reversal invariance) 
requires that
\be
\cM^{I=0} (\pz , \rz ) &=& \;\; \cM^{I=0} (-\pz , -\rz ), 
\nonumber \\
\cM^{I=1} (\pz , \rz ) &=& -\cM^{I=1} (-\pz , -\rz ) ,
\label{symmetry}
\ee
for same $t$; the symmetry of $\cM^G$ is the same as that
of $\cM^{I=0}$. In fact, using in addition hermitean conjugation one 
obtains a stronger symmetry relating the functions with 
$\rz \rightarrow -\rz$ and same $\pz$ \cite{MPW97}:
\be
\cM^{I=0, 1} (\pz , \rz ) &=& \cM^{I=0, 1} (\pz , -\rz );
\label{muenchen_me}
\ee
this will be discussed in detail in Subsection \ref{subsec_trouble}
\par
{\it Skewed distributions.}\ 
In principle, the matrix elements Eqs.(\ref{me_def}), (\ref{me_gluon}) 
can be considered as functions of the invariants $\pz$ and $\rz$ as 
independent variables, defined in the physical region. However, in 
the amplitude for hard processes such as DVCS off 
the pion the matrix elements enter with some fixed ratio of $\rz$ and 
$\pz$,
\be
\rz &=& \xi \, \pz ,
\label{skewedness}
\ee
which is dictated by the kinematics of the process; for instance,
in DVCS $\xi$ is related to the Bjorken variable 
($-1 < \xi / 2 < 1$) \cite{Ji1,Ji2,Rad1,Rad2}. This suggests 
to define a ``one--dimensional'' 
spectral representation of the matrix elements in the form
\be
\cM^{I=0,1} (\pz, \,\, \rz = \xi \pz; \,\, t ) 
&=& 2 \pz \, \int_{-1}^1 dX \;
e^{-i X \pz} \; H^{I=0,1} (X, \xi ; t) ,
\label{skewed_def}
\ee
where $H^{I=0,1} (X, \xi ; t)$ are called the skewed quark distributions
in the pion (the definition of the gluon distribution is analogous). 
The limits $\pm 1$ for the integral over the
parameter $X$ follow from rather general 
considerations \cite{Ji2,Rad1,Rad2}.
One can also give an explicit expression for $H$: Introducing a 
dimensionless light--like vector, $n$, and setting $z = \tau n$, one can 
invert Eq.(\ref{skewed_def}) and obtains \cite{Ji1,PPPBGW97}
\be
\lefteqn{ H (X, \xi; t) \;\; = \;\; 
\frac{1}{2} \int\frac{d\tau}{2\pi} e^{i\tau X \, p \cdot n} } &&
\nonumber \\[1ex] 
&& \times \langle \pi (p - r/2 ) | \; \bar\psi (-\tau n/2 ) \,\,
\hat{n} \,\, [-\tau n/2, \tau n/2] \, \psi (\tau n/2)
\; | \pi (p + r/2 ) \rangle 
\label{H_explicit}
\ee
[with isospin decomposition as in Eq.(\ref{me_def})], and similarly 
for the gluon distribution. The symmetry property
Eq.(\ref{symmetry}) requires $H^{I=0}$ to be an odd function of $X$, 
$H^{I=1}$ to be even. From the stronger symmetry Eq.(\ref{muenchen_me})
it follows that the skewed distribution is an even function of 
$\xi$ for any $X$;
this was first noted in the case of the nucleon in Ref.\cite{MPW97}.
\par
The skewed distributions possess a simple partonic interpretation,
the character of which depends on the relation of $X$ to the 
skewedness, $\xi$; see Refs.\cite{Rad1,Rad2,Rad3} for details. 
For $X > \xi /2$ and $X < -\xi /2$ the skewed quark distributions
describe the amplitude for emission and reabsorption of a quark/antiquark 
in the infinite--momentum frame, and thus has properties analogous to the
usual quark/antiquark distribution functions. For $-\xi /2 < X < \xi /2$,
on the other hand, they have the character of distribution amplitudes
for the creation of a quark/antiquark pair. One may thus expect
the behavior of these functions to be quite different in the
two regions.
\par
In particular, in the forward limit of the matrix element, 
$r \rightarrow 0$ and $\xi \rightarrow 0$, the skewed quark 
distributions reduce to the usual quark/antiquark distributions 
in the pion:
\be
H^{I=0} (X, \, \xi = 0; \, t = 0) &=& \frac{1}{2}
\left[ \theta (X) \, q_{\rm s} (X) \; - \; 
\theta (-X) \, q_{\rm s} (-X) \right] ,
\nonumber \\[1ex]
H^{I=1} (X, \, \xi = 0; \, t = 0) &=& 
\;\;\; \theta (X) q_{\rm v} (X) \; + \; \theta (-X) q_{\rm v} (-X) ,
\label{distribution_def}
\ee
where $q_{\rm s} (X), q_{\rm v} (X)$ correspond, respectively, to the 
singlet (quark plus antiquark) and valence (quark minus antiquark) 
distributions in a physical pion:
\be
q_{\rm s} (X) &=& [u + \bar u]_{\pi^\pm}(X) \;\; = \;\;
[d + \bar d]_{\pi^\pm}(X) 
\; = \; [u + \bar u]_{\pi^0}(X) \;\; = \;\;
[d + \bar d]_{\pi^0}(X),
\nonumber \\[1ex] 
q_{\rm v} (X) &=& \pm [u - \bar u]_{\pi^\pm}(X) \;\; = \;\;
\mp [d - \bar d]_{\pi^\pm}(X) .
\ee
\par
The moments of the skewed distribution, Eq.(\ref{H_explicit}), are
given by non-forward matrix elements of local twist--2 spin--$N$ 
operators in the pion, which are parametrized by generalized form
factors. On general grounds, the non-forward matrix elements of the 
spin--$N$ operators are irreducible rank--$N$ tensors constructed from 
the momenta $p$ and $r$, so the moments of Eq.(\ref{H_explicit}) are 
polynomials of degree at most $N$ in $\xi$ \cite{Rad2,Rad3,JiReview}.
In particular, the second moment of the isoscalar skewed 
distribution is related to the form factor of the QCD 
energy--momentum tensor \cite{Ji2}. For the pion
this form factor at $t = 0$ can be computed from first principles 
using the chiral Lagrangian (see Appendix \ref{app_em}), and one 
obtains a generalized momentum sum rule for the pion,
\be
\int_{-1}^1 dX \, X \, \left[ H^{I=0} + H^G \right] (X, \xi; t = 0) &=& 
\frac{1}{2} \left( 1 \; - \; \frac{\xi^2}{4} \right) .
\label{em_sum_rule}
\ee
The isovector skewed distribution in the pion is normalized to the
pion electromagnetic form factor. For any $t < 0$:
\be
\int_{-1}^1 dX  \, H^{I=1}(X,\xi; t) &=& F_\pi^{{\rm e.m.}}(t) .
\label{norm_isovector}
\ee
\par
\begin{figure}
\setlength{\epsfxsize}{15cm}
\setlength{\epsfysize}{15cm}
\epsffile{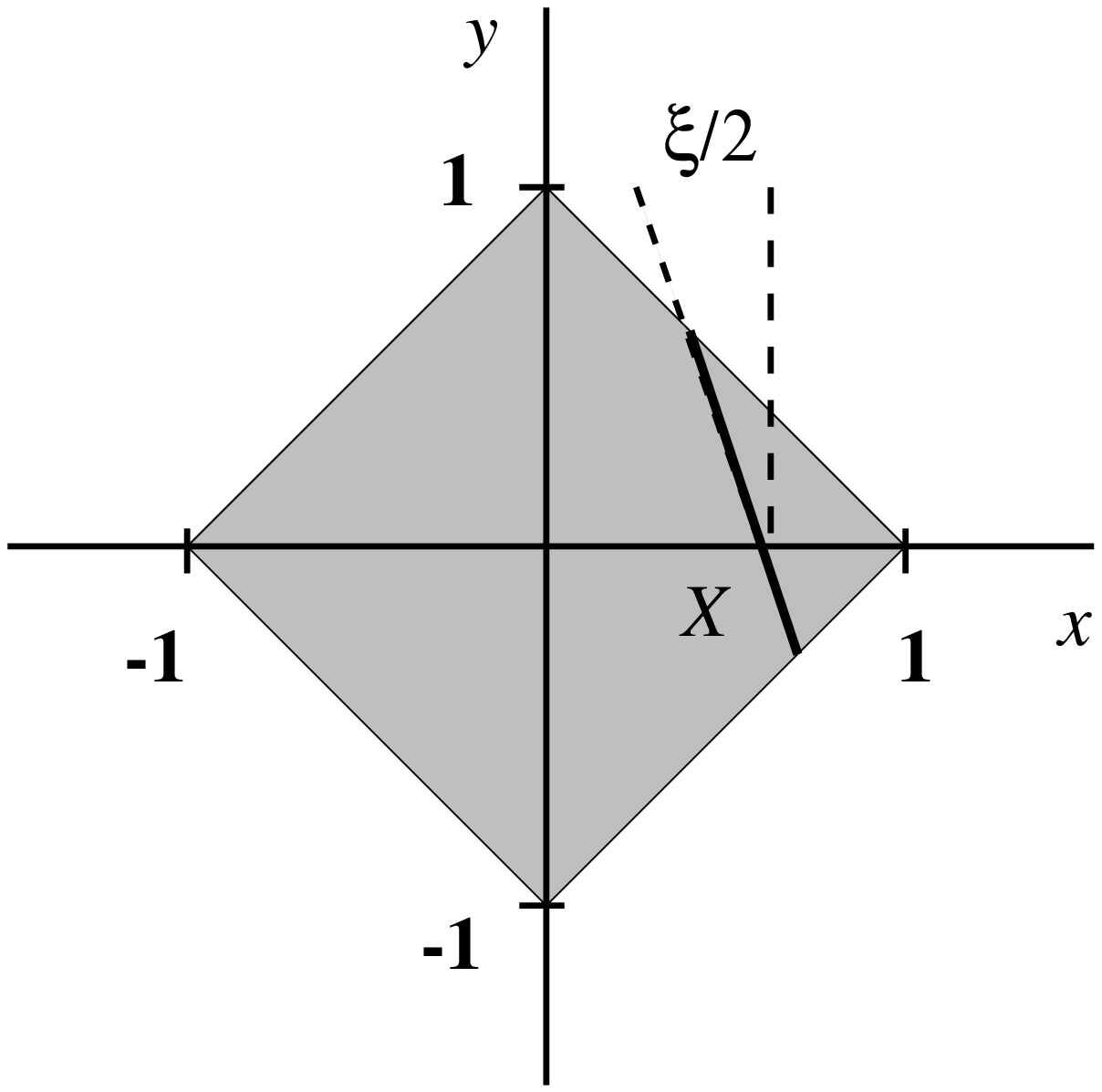}
\caption[]
{The range of the variables $x$ and $y$ in the double distribution,
Eq.(\ref{double_naive}). The reduction to the skewed distribution,
$H(X, \xi)$, is achieved by integrating the double distribution over 
the line $x + y \xi /2 = X$, {\it cf.}\ Eq.(\ref{reduction_naive}), 
shown here for the case that $X > \xi /2$ ({\it thick line}).}
\label{fig_rhombus}
\end{figure}

{\it Double distributions.}\ 
Alternatively to the skewed distribution, Eq.(\ref{skewed_def}), one 
can try to formulate a ``two--dimensional'' spectral
representation of the matrix element Eq.(\ref{me_def}),
as a function of $\rz$ and $\pz$ as independent variables.
In the spirit of Refs.\cite{Rad1,Rad2} we could
write for the pion matrix elements a spectral representation in terms 
of a single function of two variables in the form
\be
\cM^{I=0, 1} (\pz, \,\, \rz ; \,\, t ) &=& 2 \pz \,
\int_{-1}^1 dx \; e^{-i x \pz} \;
\int_{-(1 - |x|)}^{1 - |x|} dy \; e^{-i y \rz / 2} \;
F^{I=0, 1}(x, y ; t) ,
\label{double_naive}
\ee
where the functions $F^{I=0, 1}$ are called double 
distributions\footnote{We consider
here the ``modified'' double distribution of Ref.\cite{Rad2}, which
is appropriate for the symmetric choice of the momenta of the
incoming and outgoing pion in Eq.(\ref{me_def}).}. Here the range
of the variables $x, y$ is limited to \cite{Rad1,Rad2}
\[
-1 \; \leq x \; \leq 1, \hspace{2cm} -(1 - |x|) \; \leq y
\; \leq 1 - |x| ,
\]
see Fig.\ref{fig_rhombus}. The property Eq.(\ref{muenchen_me}) implies 
that \cite{MPW97}
\be
F^{I=0, 1} (x, y; t) &=& F^{I=0, 1} (x, -y; t) .
\label{muenchen_double}
\ee
The skewed distribution, Eq.(\ref{skewed_def}), is obtained
as a one--dimensional ``section'' of this two--variable
function, imposing a particular ``skewedness'', $\xi$:
\be
\int_{-1}^{1} dx \int_{-(1 - |x|)}^{1 - |x|} dy \;
\delta (X - x - y\xi / 2) \, F^{I=0, 1}(x, y; t) &=& 
H^{I=0, 1}(X, \xi; t) .
\label{reduction_naive}
\ee
This reduction process is illustrated in Fig.\ref{fig_rhombus}.
In particular, in the forward limit the usual quark/antiquark 
distribution is recovered as
\be
\int_{-(1 - |X|)}^{1 - |X|} dy \; F^{I=0}(X, y ; t = 0) &=& 
\frac{1}{2}
\left[ \theta (X) \, q_{\rm s} (X) \; - \; 
\theta (-X) \, q_{\rm s} (-X) \right] 
\label{forward_naive}
\ee
and similarly for the isovector component, 
{\it cf.}\ Eq.(\ref{distribution_def}).
\par
The main reason for interest in a double distribution representation is 
the possibility to relate skewed distributions with different values of
$\xi$. With certain assumptions about the behavior of the 
double distribution a number of statements about the skewed
distributions follow immediately from the reduction 
formula, Eq.(\ref{reduction_naive}); see Refs.\cite{Rad1,Rad2} for an 
extensive discussion. For instance, if the double distribution were
continuous everywhere on its region of support 
(see Fig.\ref{fig_rhombus}), the skewed distribution
would be a continuous function of $x$ and $\xi$. The double distribution
is also convenient for model building, since any model of the double 
distribution, when inserted in the reduction formula, produces skewed 
distributions satisfying the polynomiality condition for the
moments \cite{Rad1,Rad2,MPW97,Rad3}. However, in order to be
practically relevant, such applications require
understanding of the general behavior of the double distributions, 
in particular, of their possible singularities.
\subsection{Trouble with double distributions}
\label{subsec_trouble}
When discussing properties of double distributions (such as their 
singularities) one should keep in mind that the behavior 
of these functions is determined by the behavior of the matrix 
element, Eq.(\ref{me_def}), as a function of $\pz$ and $\rz$, as well 
as by the particular way in which one writes the spectral representation 
for it. This concerns, in particular, the number of independent 
``twist--2 structures'' one takes into account in 
the decomposition of the matrix element.
We shall argue now that it is not always adequate to represent 
the pion non-forward matrix element in the form of 
Eq.(\ref{double_naive}), as a double spectral integral
with a single prefactor, $\pz$. This form is incompatible with
general features of the dependence of the isoscalar matrix element, 
Eq.(\ref{me_def}), on $\pz$ and $\rz$, and insisting on it
one would incur severe singularities in the double distribution.
\par
In order to obtain information about the behavior of the function 
$\cM (\pz , \rz , t)$, Eq.(\ref{me_def}), it is useful to consider 
instead of Eq.(\ref{me_def}) 
the more general matrix element of the non-local vector
operator [isospin components are defined in analogy to Eq.(\ref{me_def})]
\be
\langle \pi (p - r/2 ) | \; \bar\psi (-z/2 ) 
\gamma_\mu \, [-z/2, z/2] \, \psi (z/2)
\; | \pi (p + r/2 ) \rangle &\equiv& \cM_\mu (p, r, z) ,
\ee
from which the matrix element Eq.(\ref{me_def}) is
obtained by contraction with the light--cone vector, $z^\mu$.
On general grounds, this matrix element can be parametrized as
(for both $I = 0$ and $1$)
\be
\cM_\mu (p, r, z)
&=& 2 p_\mu G \; + \; r_\mu G_\parallel \; + \; 
\mbox{term $\propto z_\mu$} ,
\label{vector_operator}
\ee
where $G$ and $G_\parallel$ are generalized form factors depending on 
$\pz , \rz$ and 
$t$. The terms proportional to $z_\mu$ vanish upon contraction with 
$z^\mu$ and does not contribute to the twist--2 part matrix element 
Eq.(\ref{me_def}). The term proportional to $r_\mu$, however, does 
contribute to Eq.(\ref{me_def}). In fact, it is the presence of this
structure which causes trouble in the double--distribution 
representation of the isoscalar matrix element, Eq.(\ref{double_naive}).
\par
In the limit $z \rightarrow 0$ the operator in Eq.(\ref{vector_operator})
reduces to the local vector current, which is conserved. This
implies that $G_\parallel (z = 0) \equiv 0$ for all $t$, {\it i.e.}, the 
matrix element is ``transverse'' ($\propto p_\mu$). However, the non-local 
operator with $z \neq 0$ is generally not 
conserved, so there is no reason for $G_\parallel$ to be zero for 
$z \neq 0$.
Actually, current conservation is only a sufficient
condition for $G_\parallel$ to be zero, not a necessary one. For
$z = 0$ one obtains $G_\parallel \equiv 0$ already from time 
reversal invariance and the hermiticity of the local current 
operator. Applying the same symmetry transformations 
to the non-local operator, one finds that for the isoscalar
matrix element
\be
G^{I=0} (\pz , \,\, \rz , \,\, t) &=& \;\;\; 
G^{I=0} (\pz , \,\, -\rz , \,\, t) , 
\nonumber \\
G_\parallel^{I=0} (\pz , \,\, \rz , \,\, t) &=& 
-G_\parallel^{I=0} (\pz , \,\, -\rz , \,\, t) .
\label{muenchen_form}
\ee
In the local case $\rz = 0$, and $G^{I=0}_\parallel$ would be zero 
identically in $t$. However, in the general case, $z \neq 0$, there is 
again no reason for $G_\parallel^{I=0}$ to be zero.
\par
The presence of a ``longitudinal'' ($\propto r_\mu$) part of the 
vector matrix element, Eq.(\ref{vector_operator}), means that the
matrix element $\cM^{I=0}$, obtained by contracting 
Eq.(\ref{vector_operator}) with $z^\mu$, contains in addition to
the $\pz$--term a piece with prefactor $\rz$,
\be
\cM^{I=0} (\pz , \rz , t) &=& \pz \, G^{I=0} (\pz , \rz , t) 
\; + \; \rz \, G_\parallel^{I=0} (\pz , \rz , t) .
\label{M_from_vector}
\ee
In particular, since generally $G_\parallel^{I=0} \neq 0$, $\cM$ 
does not vanish in the limit $\pz \rightarrow 0$ and $\rz \neq 0$:
\be
\cM^{I=0} (\pz \rightarrow 0, \rz \neq 0, t) &\neq& 0 .
\label{M_nonzero}
\ee
One can easily see that this implies that the matrix element
$\cM^{I=0}$ cannot be represented in the form Eq.(\ref{double_naive})
with a non-singular double distribution $F^{I=0}(x, y)$.
Suppose $F^{I=0} (x, y)$ were non-singular in its range of support. 
It would then define the matrix element on the 
L.H.S.\ as an analytic function of the variables $\pz$ and $\rz$, which
could be continued to a point (in the 
unphysical region) where $\pz = 0$, but $\rz \neq 0$. At this point
the R.H.S.\ of Eq.(\ref{double_naive}) vanishes because of the
prefactor $\pz$, but not the matrix element on the L.H.S., {\it cf.}\
Eq.(\ref{M_nonzero}). Clearly, this implies that $F^{I=0}(x, y)$ must be 
singular, in one way or another.
\par
What would be the character of these singularities? Trying to
absorb a $\pz$--independent piece in the integral Eq.(\ref{double_naive})
would amount to finding an integral representation of $1/\pz$ in the form
\be
\frac{1}{\pz} &=& \int_{-1}^1 dx \; e^{-i x \pz} \; f(x) ,
\label{singularity}
\ee
with $f(x)$ some generalized function. Assuming that the integral on the 
R.H.S.\ can be continued to $\pz \rightarrow 0$, one concludes that no 
Mellin moments of the function $f(x)$ exist. In particular, this means that 
the singularity in $f(x)$ cannot be of delta--function type.
(We shall return to this point below.)
Thus, we conclude that, although the two contributions to $\cM^{I=0}$ 
in Eq.(\ref{M_from_vector}) are not structurally distinct, it is not
possible to include the $\rz$--term in a double distribution
representation of the form Eq.(\ref{double_naive}) staying
within the usual class of generalized functions.
\par
It is important to mention that the difficulties noted here
do not concern the representation of the matrix element 
through a skewed distribution, Eq.(\ref{skewed_def}).
In this case one writes the representation of the matrix element 
under the condition $\rz = \xi \pz$, with $\xi$ fixed.
In particular, repeating the above argument 
and taking in Eq.(\ref{skewed_def})
the limit $\pz \rightarrow 0$ we would now also
have $\rz \rightarrow 0$, so that both 
L.H.S.\ and R.H.S.\ of Eq.(\ref{skewed_def}) would vanish,
and there is no need for $H^{I=0}$ to be singular.
\par
Since there are many advantages in a double--distribution 
representation of the non-forward matrix elements, it is 
worthwhile to think how Eq.(\ref{double_naive}) could
be modified to allow for a double spectral representation in terms of 
standard generalized function. The origin of the problems
with the form Eq.(\ref{double_naive}) is that the
matrix element does not go to zero in the limit $\pz \rightarrow 0$, 
Eq.(\ref{M_nonzero}). One possibility would be to simply omit the
prefactor $\pz$ in Eq.(\ref{double_naive}); however, this would
result in a function $F^{I=0}(x, y)$ which does not reduce to the
usual parton distribution in the forward limit, $r\rightarrow 0$,
{\it cf.}\ Eq.(\ref{forward_naive}), and would not be useful
for model building. Alternatively, one could add to 
Eq.(\ref{double_naive}) a term not vanishing in the
limit $\pz \rightarrow 0$. Minimally, this could be a term
depending only on $\rz$, which can be represented by a 
one--dimensional spectral integral:
\be
\cM^{I=0} (\pz, \,\, \rz ; t) &=& 2 \pz \, 
\int_{-1}^1 dx \; e^{-i x \pz} \; 
\int_{-(1 - |x|)}^{1 - |x|} dy \; e^{-i y \rz / 2} \; 
F^{I=0} (x, y; t) 
\nonumber \\
&&+ \;\; \rz \, \int_{-1}^{1} dy \; e^{-i y \rz / 2} \; 
D (y; t) .
\label{double_minimal}
\ee
The functions $F^{I=0}(x, y; t)$ and $D(y; t)$ are uniquely defined 
if we understand the first term to be a representation of 
$\cM^{I=0} (\pz , \rz; t) - \cM^{I=0} (0, \rz ; t)$, 
the second of $\cM^{I=0} (0, \rz ; t)$, {\it i.e.}, as a 
``subtraction term''. The explicit factor $\rz$ in front of 
the second term is natural since $\cM^{I=0} (0, \rz \rightarrow 0) = 0$.
The support of $D$ is limited to $-1 \leq y \leq 1$, {\it i.e.},
this function has the character of a distribution amplitude.
Time reversal and hermiticity, Eq.(\ref{muenchen_form}), require that
(same $t$)
\be
F^{I=0} (x, y) &=& F^{I=0} (x, -y) ,
\hspace{2cm} D (y) \;\; = \;\; -D (-y) ,
\label{muenchen_minimal}
\ee
{\it i.e.}, the behavior with respect to $y \rightarrow -y$ of the 
new function $D(y)$ is opposite to that of the usual double 
distribution, Eq.(\ref{muenchen_double}).
\par
The skewed distribution which follows from the new representation
Eq.(\ref{double_minimal}) is now the sum of two contributions:
\be
H^{I=0} (X, \xi ; t) &=& 
\int_{-1}^{1} dx \int_{-(1 - |x|)}^{1 - |x|} dy \;
\delta (X - x - y\xi / 2) \; F^{I=0}(x, y; t) 
\nonumber \\
&& \; + \;\; \mbox{sign} (\xi ) \, 
D \left( \frac{2 X}{\xi} ; t \right) .
\label{reduction_minimal}
\ee
Note that both contributions are even functions of $\xi$, in accordance
with the general symmetry of the skewed distribution following
from Eq.(\ref{muenchen_me}). The first piece follows the usual reduction 
formula, Eq.(\ref{reduction_naive}), and is generally non-zero in the 
entire range $-1 < X < 1$. The second piece is obtained
by substituting in Eq.(\ref{double_minimal})
$\rz = \xi \pz$ and changing the integration variable to $\xi y/2$. 
Since the support of $D$ is limited to $-1 \leq y \leq 1$ this 
contribution to the skewed distribution is present only for 
$-\xi / 2 \leq X \leq \xi /2$.
Thus, the need to include the ``subtraction term'' in the
double distribution representation naturally leads to a 
skewed distribution with essentially different behavior in the
regions $|X| < \xi / 2$ and in $|X| > \xi / 2$.
\par
So far we have explored the consequences of the presence
of ``longitudinal'' terms in the matrix element 
Eq.(\ref{vector_operator}), or of property 
Eq.(\ref{M_nonzero}), from a general point of view, 
arguing that there is no reason
for such contribution to be zero. In Section \ref{sec_model} we
perform a model calculation of the pion matrix elements at a 
low normalization point in the effective chiral theory based on
the instanton vacuum, wich shows that such terms in the matrix 
element do indeed appear, and lead to the ``two--component''
form of the skewed distribution described above.
\par
The region $\pz \rightarrow 0$ and $\rz \neq 0$ implied in the
limit in Eq.(\ref{M_nonzero}) corresponds to values of $|\xi | > 2$, 
which are not physically accessible in DVCS. However, using crossing 
invariance one can relate the function $\cM (\pz , \rz )$ in the 
unphysical region, Eq.(\ref{M_nonzero}), to the matrix element 
for two--pion production by a light--ray operator in the {\it physical} 
region (see Section \ref{sec_crossing}). The latter can be measured 
({\it e.g.}\ in $\gamma^\ast \gamma \rightarrow \pi\pi$ 
and $\gamma^\ast N \rightarrow \pi\pi N$ reactions)
and is generally non-zero, providing additional evidence for 
the presence of $\rz$--terms and the property Eq.(\ref{M_nonzero}).
\par
A simple dynamical explanation for the origin of $\rz$--terms in the 
isoscalar pion matrix element can be found by considering 
``resonance exchange'' contributions to the matrix element, in the
spirit of the vector dominance model for the pion electromagnetic
form factor \cite{Sakurai60}. By this we mean ``factorized''
contributions to the matrix element in which the pion and the
light--ray operator communicate by $t$--channel exchange of a resonance
characterized by a twist--2 distribution amplitude, as are shown 
schematically in Fig.\ref{fig_exchange}. In Appendix \ref{app_resonance} 
we derive a general formula describing the contribution resulting from 
the exchange of a resonance of arbitrary spin to the pion matrix
element. In particular, we show there that the property 
Eq.(\ref{M_nonzero}) of the isoscalar matrix element is naturally
obtained from exchange of even--spin isoscalar resonances.
Far from being a complete dynamical description of the
non-forward matrix element, this phenomenological model helps
to develop an intuitive understanding why the structures
described above appear.
\par
The double distribution $F^{I=0}(x, y)$ in Eq.(\ref{double_minimal})
is a generalized function which may contain delta function type
singularities. Such terms in the double distribution
[in the restricted ansatz Eq.(\ref{double_naive})] were studied 
by Radyushkin in connection with resonance exchange contributions 
to the non-forward nucleon matrix elements \cite{Rad2,Rad3}. 
We already argued above that the pure $\rz$--terms in the isoscalar 
pion matrix element, {\it cf.}\ Eq.(\ref{M_nonzero}), cannot be 
described by delta function type contributions 
to $F^{I=0}(x, y)$. It is interesting to verify this at the level 
of the reduction formula, 
Eq.(\ref{reduction_minimal}). Could delta function 
contributions to $F^{I=0}(x, y)$ mock up the 
structure of the $D$--term in $H^{I=0}$? A term in 
$F^{I=0} (x, y)$ of the form 
$\delta (x) \phi (y)$ would give a contribution to the
skewed distribution $H^{I=0} \propto \phi (2 X / \xi )\, /|\xi |$,
which is ruled out because $H^{I=0}$ must at the same time be odd 
in $X$ and even in $\xi$. One thus has to turn to derivatives
of $\delta (x)$. A term $\delta' (x) \chi (y)$ would give 
a contribution 
$H^{I=0} \propto \mbox{sign} (\xi ) \, \chi' (2 X / \xi ) \, /\xi^2$,
which can be non-zero but cannot describe the contribution
generated by $D(y)$ in Eq.(\ref{reduction_minimal}) (consider
for example the forward limit). This argument can easily be 
extended to any derivative of $\delta (x)$. Thus, we conclude that 
the term generated by $D(y)$ in Eq.(\ref{double_minimal}) represents
a genuine separate structure which cannot be obtained from
delta function contributions to $F^{I=0}(x, y)$.\footnote{It is 
amusing to note that, rather than a derivative of $\delta (x)$ this 
term represents, in a sense, 
an ``integral'' of $\delta (x)$, {\it cf.}\ Eq.(\ref{singularity}).}
\par
We remark that in the resonance exchange model of 
Appendix \ref{app_resonance} exchange of even--spin resonances 
generally contributes to both $D(y)$ and to delta function terms in 
$F^{I=0}(x, y)$, {\it cf.}\ Eq.(\ref{R_M}). Spin--0
(``sigma meson'') exchange is special in that it contributes 
only to $D(y)$.
\par
In the amplitude for hard exclusive processes such as DVCS
the skewed distribution is convoluted with a hard scattering
kernel which is singular at $X = \pm \xi /2$ 
\cite{Ji1,Ji2,Rad1,Rad2}. For this integral to exist ({\it i.e.}, 
for factorization to hold) it is important that the skewed 
distribution be continuous in $X$ at these points. Assuming that
the first term on the R.H.S.\ of Eq.(\ref{reduction_minimal})
is continuous, this would be satisfied if
\be
D(y; t) &\rightarrow& 0 \hspace{1.5cm} (y \rightarrow \pm 1).
\label{endpoint}
\ee
We shall see below that it is indeed reasonable to expect that
$D(y; t)$ satisfies this property, reminiscent of a meson
distribution amplitude. Model calculations of the matrix elements
in the effective chiral theory bases on the instanton vacuum and 
in a ``resonance exchange'' model give rise to functions
satisfying Eq.(\ref{endpoint}).
\par
The representation Eq.(\ref{double_minimal}) is the
minimal modification of Eq.(\ref{double_naive}) consistent 
with Eq.(\ref{M_nonzero}). For some
purposes ({\it e.g.}\ crossing symmetry) it could be convenient 
to have a representation which is symmetric with respect to $\pz$ 
and $\rz$. One could write:
\be
\cM^{I=0} (\pz, \,\, \rz ) &=& 2 \pz \, 
\int_{-1}^1 dx \; e^{-i x \pz} \; 
\int_{-(1 - |x|)}^{1 - |x|} dy \; e^{-i y \rz / 2} \; 
F^{I=0} (x, y ) 
\nonumber \\
&&+ \;\; \rz \, \int_{-1}^1 dx \; e^{-i x \pz} \; 
\int_{-(1 - |x|)}^{1 - |x|} dy \; e^{-i y \rz / 2} \; 
F_\parallel^{I=0} (x, y ) .
\label{double_two}
\ee
The two functions $F^{I=0}$ and $F^{I=0}_\parallel$ 
would be uniquely 
determined if we defined them as the spectral representation of the 
generalized form factors $G^{I=0}(\pz , \rz , t)$ and 
$G^{I=0}_\parallel(\pz , \rz , t)$ of the vector operator, 
Eq.(\ref{vector_operator}).\footnote{Strictly speaking, 
we have no general proof that a double spectral representation for the
form factors $G$ and $G_\parallel$ exists. At least in our model 
calculations in Section \ref{sec_model} and Appendix \ref{app_resonance} 
we shall encounter only contributions to the matrix element which can
be represented by Eq.(\ref{double_two}) with 
$F, F_\parallel$ having
at most delta function singularities.} Again, Eq.(\ref{muenchen_form}) 
requires that
\be
F^{I=0} (x, y ) &=&  F^{I=0} (x, -y ) , 
\hspace{2cm}
F_\parallel^{I=0} (x, y ) \;\; = \;\; 
-F_\parallel^{I=0} (x, -y ) . 
\label{bochum}
\ee
In terms of these new functions the skewed distribution 
would now be given by the reduction formula 
Eq.(\ref{reduction_naive}) with 
\be
F(x, y) &\rightarrow& F (x, y) \; + \; 
\frac{\xi}{2} \, F_\parallel (x, y) .
\ee
Note that both $F$ and $F_\parallel$ are generalized functions 
which may contain delta--function singularities.
\par
Finally, let us note that for the isovector pion matrix element,
$\cM^{I=1}$, the original form of the double distribution 
representation, Eq.(\ref{double_naive}), works fine, and no
``subtraction terms'' of the kind in Eq.(\ref{double_minimal}) 
are required. This is because $\cM^{I=1}$ is odd in $\pz$
for any $\rz$ and t [as follows from combining Eq.(\ref{symmetry}) 
and Eq.(\ref{muenchen_me})], and thus 
\be
\cM^{I=1} (\pz \rightarrow 0, \rz \neq 0, t) &\rightarrow& 0 .
\ee
In the resonance exchange model this is again easily understood;
it follows from the fact that isovector two--pion resonances have odd 
spin, see Eq.(\ref{R_M}) in Appendix \ref{app_resonance}.
\subsection{The nucleon}
\label{subsec_nucleon}
We now turn to non-forward matrix elements in the nucleon. By a simple
extension of the arguments offered in the previous subsection for the
pion, we show that also in the case of the nucleon the double distribution
representation of Refs.\cite{Rad1,Rad2,Rad3} should be modified to take 
into account all possible twist--2 structures.
\par
The object of interest now is the nucleon matrix element of the
twist--2 light ray operator of Eq.(\ref{me_def}). Again we 
distinguish the isoscalar and isovector matrix 
elements:
\be
\lefteqn{
\langle N (p - r/2 ) , T_3' , \lambda' | \; \bar\psi (-z/2 ) 
\left\{ \begin{array}{c} 1 \\[2ex] \tau^c \end{array} \right\}
\hat{z} \, [-z/2, z/2] \, \psi (z/2)
\; | N (p + r/2 ) , T_3 , \lambda \rangle } 
\hspace{3cm} &&
\nonumber \\
&=& \left\{ \begin{array}{r}
2 \delta_{T_3 T_3'} \; \cM^{I=0}
({\lambda' , \lambda}; \pz , \,\, \rz ; \,\, t)
\\[2ex]
2 \left( \tau^{c} \right)_{T_3 T_3'} \cM^{I=1}
({\lambda' , \lambda} ; \pz , \,\, \rz ; \,\, t) ,
\end{array} \;\;\;\;\;
\right.
\label{me_def_nucleon}
\ee
where $T_3 , T_3'$ denote the isospin projection 
($T_3 , T_3' = \pm 1/2$ for proton/neutron). The only difference to the
pion is that now the functions $\cM^{I=0, 1}$ depend also on the 
helicities of the incoming ($\lambda$) and outgoing ($\lambda'$) 
nucleon. In analogy to the pion, Eq.(\ref{vector_operator}), let us 
consider also the matrix element of the more basic light--ray operator 
with $\gamma_\mu$ [the isospin decomposition is analogous to 
Eq.(\ref{me_def_nucleon}) and not written],
\be
\langle N (p - r/2 ) , \lambda' | \; 
\bar\psi (-z/2 ) \gamma_\mu \, [-z/2, z/2] \, \psi (z/2)
\; | N (p + r/2 ) , \lambda \rangle
&\equiv& \cM_\mu (\lambda' , \lambda; p, r, z) ,
\nonumber \\
\ee
which on general grounds can be parametrized as (for both $I = 0$ and $1$)
\be
\cM_\mu &=& 
\bar U' \left[ \gamma_\mu \; G_1 \; + \; \frac{i}{2M_N} 
\sigma_{\mu\nu} r^\nu G_2 \; + \; r_\mu G_\parallel \; + \; 
\ldots \right] U 
\label{vector_operator_nucleon}
\ee
where $\bar U' \equiv \bar U (p - r/2 , \lambda'),
\;\; U \equiv U (p + r/2 , \lambda )$ are the nucleon spinors.
We have not written explicitly terms which vanish upon contraction
with $z^\mu$ and do not contribute to the twist--2 part
(such as $z_\mu, \, \sigma_{\mu\nu} z^\nu$).
Here $G_1, G_2, G_\parallel$ are generalized form factors depending on
$\pz, \rz$ and $t$. In the limit of a local operator, 
$z \rightarrow 0$, $G_1$ and $G_2$ reduce to the usual Dirac form 
factors for the vector current, and $G_\parallel \rightarrow 0$ because of
current conservation. However, as in the case of the pion,
for $z \neq 0$ the term $\propto r_\mu$ is generally non-zero, 
$G_\parallel \neq 0$.
\par
Contracting Eq.(\ref{vector_operator_nucleon}) with $z^\mu$ we 
obtain the twist--2 matrix element, Eq.(\ref{me_def_nucleon})
(for both $I = 0$ and $1$):
\be
\cM (\lambda' , \lambda ; \pz , \,\, \rz ; t)
&=& \bar U' \left[ \hat{z} \; G_1 
\; + \; \frac{i}{2M_N} 
\sigma_{\mu\nu} z^\mu r^\nu G_2
\; + \;
\rz G_\parallel \right] U .
\label{M_from_vector_nucleon}
\ee
Actually, here the three structures are not independent. Using the
well--known identity
\be
\bar U' U p_\mu &=& 2 M_N \bar U' \gamma_\mu U 
\; + \; i \bar U' \sigma_{\mu\nu} U r^\nu ,
\ee
we could rewrite the third term as a linear combination
of the first two. In this way we would arrive at ($I = 0, 1$):
\be
\cM (\lambda' , \lambda ; \pz , \,\, \rz ; t)
&=& \bar U' \left[ \hat{z} \; \widetilde{G}_1 
\; + \; \frac{i}{2M_N} 
\sigma_{\mu\nu} z^\mu r^\nu \widetilde{G}_2 \right] U ,
\label{M_from_vector_alt}
\ee
where
\be
\widetilde{G}_1 &=& G_1 \; + \; 2\frac{\rz}{\pz} \, G_\parallel ,
\hspace{2cm} 
\widetilde{G}_2 \;\; = \;\; G_2 \; + \; 2\frac{\rz}{\pz} \, G_\parallel .
\label{G_tilde}
\ee
A decomposition of the form Eq.(\ref{M_from_vector_alt}) was 
assumed in Refs.\cite{Rad1,Rad2}, where a double distribution 
representation of the matrix element was proposed in the 
form
\be
\lefteqn{ \cM^{I=0, 1} (\lambda' , \lambda ; \pz , \,\, \rz ; t) }
&& \nonumber \\[1ex]
&=& \bar U' \hat{z} U \, \int_{-1}^1 dx \; e^{-i x \pz} \;
\int_{-(1 - |x|)}^{1 - |x|} dy \; e^{-i y \rz / 2} \;
F^{I=0, 1} (x, y ; t) 
\nonumber \\
&&+ \;\; \frac{i}{2 M_N} z^\mu r^\nu \bar U' \sigma_{\mu\nu} U 
\, \int_{-1}^1 dx \; e^{-i x \pz} \;
\int_{-(1 - |x|)}^{1 - |x|} dy \; e^{-i y \rz / 2} \;
K^{I=0, 1} (x, y ; t) .
\label{double_naive_nucleon}
\ee
We see that in the isoscalar case this ansatz suffers from the same 
problem as the simple ansatz for the double distribution in the 
pion, Eq.(\ref{double_naive}). Since in general $G^{I=0}_\parallel \neq 0$,
the functions $\widetilde{G}^{I=0}_1 , \widetilde{G}^{I=0}_2$ in the 
reduced decomposition, Eq.(\ref{G_tilde}), have singularities of the 
type $1/\pz$, which cannot be represented by spectral 
integrals with usual generalized functions, as described in 
the previous subsection. Thus, the conclusion is the same as
for the pion: Although the twist--2 contributions from the
``longitudinal'' part ($\propto r_\mu$) of the
vector matrix element are not structurally distinct 
from those from the ``transverse'' part ($\propto p_\mu$),
one cannot obtain them from a double distribution 
representation whose form is modeled on the ``transverse'' part.
\par
Again we stress that there is no problem with a 
representation of the matrix element Eq.(\ref{M_from_vector_alt})
in terms of skewed distributions (see Refs.\cite{Ji1,Ji2,Rad1,Rad2}
for their definition in the nucleon). In this case the factors
$\rz / \pz$ incurred in eliminating the $\rz$--term, 
Eq.(\ref{G_tilde}), are replaced by the skewedness, $\xi$,
which is a fixed external parameter.
\par
In analogy to the pion, Eq.(\ref{double_minimal}), we suggest
to modify the spectral representation Eq.(\ref{double_naive_nucleon})
by explicitly including the $\rz$--terms. A minimal variant
would be to add a term depending only on $\rz$, in which one
could absorb the $\pz$--independent part of the 
``longitudinal'' term of Eq.(\ref{M_from_vector}), 
$\rz G^{I=0}_\parallel (\pz = 0 , \rz )$:
\be
\lefteqn{ \cM^{I=0} (\lambda' , \lambda ; \pz , \,\, \rz ; t) }
&& \nonumber \\[1ex]
&=& \bar U' \hat{z} U \, \int_{-1}^1 dx \; e^{-i x \pz} \;
\int_{-(1 - |x|)}^{1 - |x|} dy \; e^{-i y \rz / 2} \;
F^{I=0} (x, y ; t) 
\nonumber \\
&&+ \;\; \frac{i}{2 M_N} z^\mu r^\nu \bar U' \sigma_{\mu\nu} U 
\, \int_{-1}^1 dx \; e^{-i x \pz} \;
\int_{-(1 - |x|)}^{1 - |x|} dy \; e^{-i y \rz / 2} \;
K^{I=0} (x, y ; t) 
\nonumber \\
&&+ \;\; \bar U' U \, \rz \, \int_{-1}^{1} dy \; e^{-i y \rz / 2} \; 
D (y ; t) .
\label{double_minimal_nucleon}
\ee
Alternatively, one could directly work with the spectral representation
of the form factors $G_1, G_2$ and $G_\parallel$, as in the representation
Eq.(\ref{double_two}) for the pion.
\par
The skewed distribution in the nucleon resulting from the full double 
distribution representation, Eq.(\ref{double_minimal_nucleon}),
again has a ``two--component'' form, since the term with $D(y)$
gives rise to a contribution non-zero only in the region
$-\xi / 2 \leq X \leq \xi /2$. (The corresponding reduction
formulas can be obtained by a trivial modification of the ones 
written in Refs.\cite{Rad1,Rad2,Rad3}.) This explains the behavior 
of the isoscalar skewed distribution, $H^{I=0}$, which was encountered 
in a model calculation in the large--$N_c$ limit \cite{PPPBGW97}.
\section{Distributions in the pion from effective chiral dynamics}
\label{sec_model}
For quantitative estimates of the non-forward
matrix elements Eq.(\ref{me_def}) and the skewed and double distributions
one has to turn to model calculations. Here we compute 
these quantities at a low normalization point in the 
low-energy effective field theory based on the instanton model of the 
QCD vacuum. This effective theory incorporates the dynamical breaking 
of chiral symmetry, and provides a realistic description of hadronic
properties of the pion and nucleon \cite{DP86,DPP88}. Its content can be
summarized in an effective action describing the interaction of a pion 
field with massive ``constituent'' quarks, in a way which is dictated 
by chiral invariance:
\be
S_{\rm eff}&=& \int d^4 x\; \bar \psi(x) \left[ i \hat{\partial}
- M F(\partial^2 ) \; e^{i \gamma_5 \tau^a \pi^a (x) / F_\pi} 
\; F(\partial^2 ) 
\right] \psi(x) .
\label{action}
\ee
Here, $\pi^a$ is the pion field, and $F_\pi = 93 \, {\rm MeV}$ is the weak 
pion decay constant. The dynamical quark mass generated in the spontaneous
breaking of chiral symmetry is momentum dependent; the form factors
$F(\partial^2 )$ are related to the instanton zero modes \cite{DP86}.
They cut loop integrals at momenta of order of the inverse average
instanton size, $\bar\rho^{-1} \approx 600\, {\rm MeV}$.
\par
The effective theory Eq.(\ref{action}) has been derived from
the instanton model of the QCD vacuum. This allows
for an unambiguous identification of the twist--2 QCD operators
with operators in the effective theory. It is understood that the 
QCD operators are normalized at a scale of the order
$\mu = \bar\rho^{-1} \approx 600\, {\rm MeV}$.
The general framework for computing parton distributions and meson
wave functions at a low normalization point within this approach has
been developed in Refs.\cite{DPPPW96,PP97}. An essential point
is that the value of the dynamical quark mass, $M$, is parametrically
small compared to the UV cutoff, $\bar\rho^{-1}$; their ratio is
proportional to the packing fraction of the instanton medium
$(M\bar\rho)^2 \sim (\bar\rho / \bar R)^4$. Qualitatively speaking
this means that in leading order in this parameter one is dealing with
structureless constituent quarks; in particular, the gluon distribution
appears only at order $(M\bar\rho)^2$. At a technical level, working in
leading order in $M\bar\rho$ means retaining only the ultraviolet divergent
part of the quark loop integrals computed with Eq.(\ref{action}), 
absorbing the ultraviolet divergence in the pion decay constant, $F_\pi$.
\par
\begin{figure}
\setlength{\epsfxsize}{12.7cm}
\setlength{\epsfysize}{7.6cm}
\epsffile{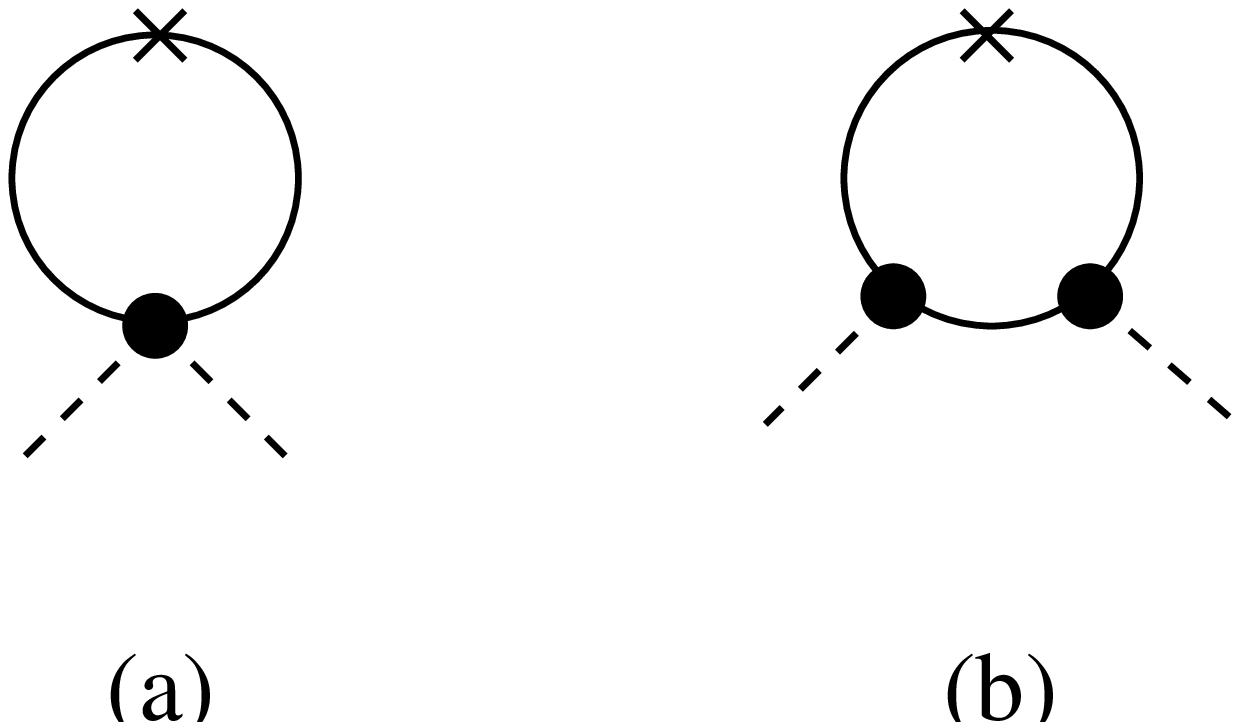}
\caption[]
{Diagrams in the effective low--energy theory contributing to the skewed 
quark distribution at a low normalization point.
The dashed line denotes the pion field, the solid line the quark
propagator with the dynamical quark mass, 
$[i\partialslash - M F^2(\partial^2 )]^{-1}$, and the filled circles
the quark--pion vertices contained in the effective action,
Eqs.(\ref{action}) and (\ref{U_gamma5}), which include a form 
factor $F(\partial^2 )$ for each quark line. Diagram (a) contributes
only to the isoscalar distribution, and vanishes in the forward
limit ($r \rightarrow 0$).}
\label{fig_diagrams}
\end{figure}

The non-linear form of the coupling of the pion to the quarks in 
Eq.(\ref{action}) is required by chiral invariance. Expanding
the exponential in powers of the pion field we obtain
\be
e^{i \gamma_5 \tau^a \pi^a (x) / F_\pi} &=&
1 \; + \; \frac{i}{F_\pi} \gamma_5 \pi^a (x) \tau^a
\; - \; \frac{1}{2F_\pi^2}\pi^a(x)\pi^a(x) \;\; + \;\; \ldots ,
\label{U_gamma5}
\ee
The effective theory contains a Yukawa--type quark--pion 
vertex as well as a two--pion quark vertex. Consequently, there
are in general two contributions to the matrix element of a twist--2
quark operator between pion states, corresponding to 
the diagrams (a) and (b) of Fig.\ref{fig_diagrams}.
The diagram (a) of Fig.\ref{fig_diagrams} contributes only to the 
flavor--singlet matrix element, while (b)
contributes both in the flavor--singlet and non--singlet case.
The Feynman integrals can straightforwardly be computed introducing 
light--cone coordinates with respect to the vector $n \sim z$; 
see Refs.\cite{PP97,PPRGW98,PW98} for details.
The integral over transverse momenta contains a logarithmic 
divergence which is cut by the form factors, $F(\partial^2 )$. 
More simply, one may keep only the logarithmically divergent
part of the diagram and absorb the logarithmic divergence in the
pion decay constant. It was shown in Refs.\cite{PPPBGW97,PPRGW98} 
that this is a legitimate approximation except in the vicinity
$X = \pm \xi / 2$. (We shall include the form factors in the
calculation later.)
\par
In this approximation the contributions of diagrams (a) and (b) 
of Fig.\ref{fig_diagrams} to the matrix elements, Eq.(\ref{me_def}), 
can be computed analytically. For the isoscalar part we obtain
(for simplicity we take $t\rightarrow 0$):
\be
\cM^{I=0} (\pz , \rz )^{\rm (a)} &=& 
2 i \, \left[ -\cos \frac{\rz}{2} + \frac{2}{\rz}
\sin\frac{\rz}{2} \right] ,
\label{M_1_unreg} \\[1ex]
\cM^{I=0} (\pz , \rz )^{\rm (b)} &=& 
2i \, \left[ \cos \pz - \frac{2}{\rz}
\sin\frac{\rz}{2} \right]
\label{M_2_unreg}
\ee
and the total result is
\be
\cM^{I=0} (\pz , \rz ) &=&
2 i \, \left[ \cos \pz - \cos\frac{\rz}{2} \right] .
\label{M_unreg}
\ee
Here contribution $(a)$ depends only on $\rz$; due to the contact nature
of the two--pion--quark vertex, Eq.(\ref{U_gamma5}), the average momentum 
$p$ does not enter in the quark loop, see Fig.\ref{fig_diagrams}. 
This contribution vanishes in the forward limit $r \rightarrow 0$.
Note that both contributions to the isoscalar matrix element, as well
as their total, behave as described in Section \ref{sec_general}: they
do not go to zero in the limit $\pz \rightarrow 0 , \; \rz \ne 0$, 
and thus cannot be represented by a double distribution in the
form Eq.(\ref{double_naive}). Within the proposed new representation, 
Eq.(\ref{double_minimal}), which allows for a $\pz$--independent part,
this model result would correspond to
\be
F^{I=0} (x, y) &=&  
\frac{1}{2} \left[ \theta (-1 < x < 0) \; - \; \theta (0 < x < 1) \right]
\,\, \delta (y) , \\[1ex]
D (y) &=& \frac{1}{2} \left[
- \theta (-1 < y < 0) \; + \; \theta (0 < y < 1) \right] ,
\label{double_minimal_model}
\ee
where $\theta (a < X < b)$ is unity if $a < X < b$ and else zero. 
That $F^{I=0}$ here is proportional to a delta function in $y$ should be 
seen as an artifact of keeping only the logarithmically divergent piece;
this would change when retaining finite terms at $t \neq 0$.
\par
The result for the isovector matrix element, Eq.(\ref{me_def}), 
at $t \rightarrow 0$ is
\be
\cM^{I=1} (\pz , \rz ) &=& \cM^{I=1} (\pz , \rz )^{\rm (b)} 
\;\; = \;\; 4 \sin\pz .
\label{M_isovec}
\ee
This matrix element vanishes in the limit $\pz \rightarrow 0$, hence
there is no problem with representing it by a double distribution 
in the form Eq.(\ref{double_naive}):
\be
F^{I=1} (x, y) &=&  \theta (-1 < x < 1) \, \delta (y) .
\label{double_isovec_model}
\ee
\par
\begin{figure}
\setlength{\epsfxsize}{15cm}
\setlength{\epsfysize}{15cm}
\epsffile{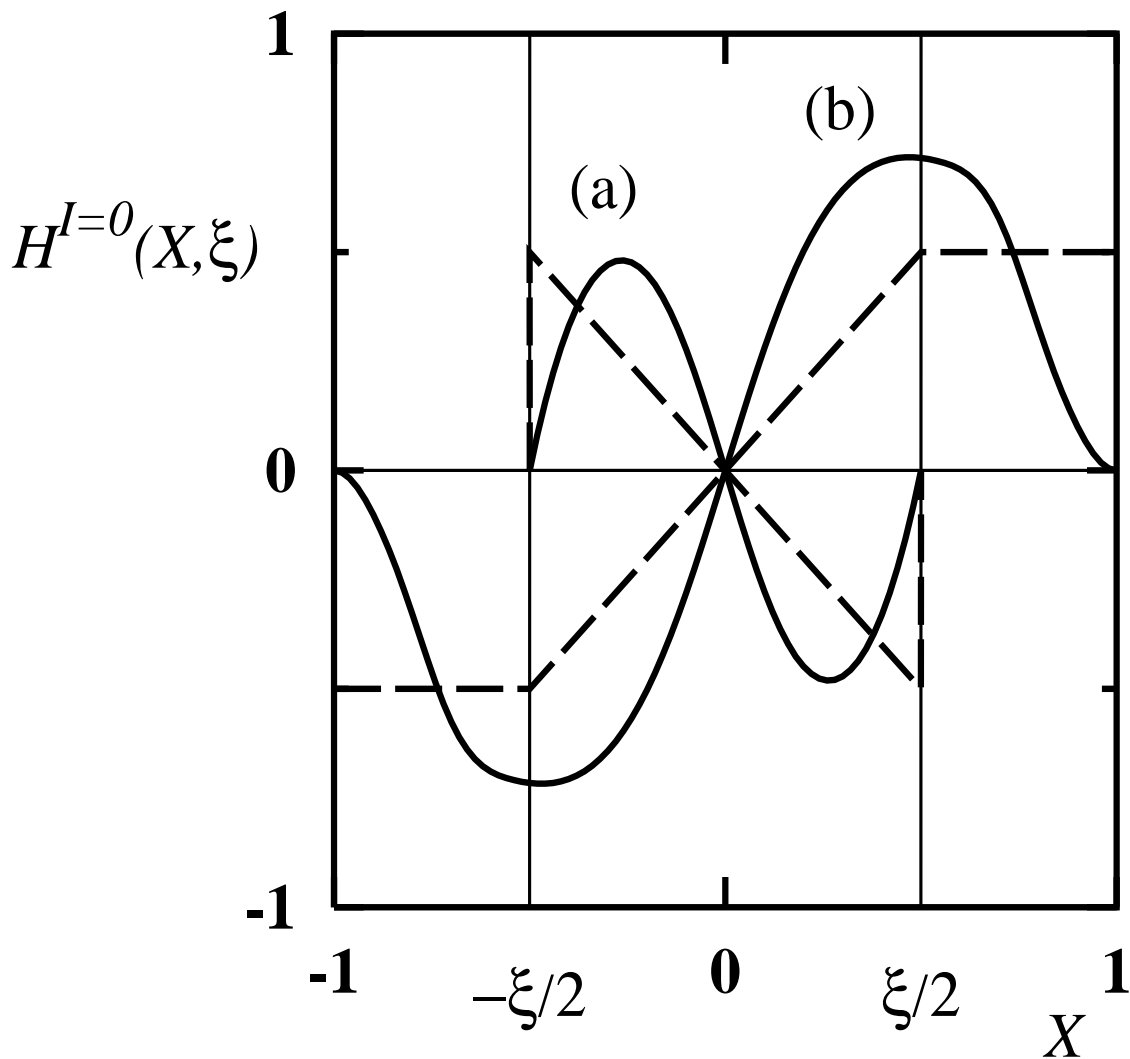}
\caption[]
{The contributions from diagrams (a) and (b) 
({\it cf.}\ Fig.\ref{fig_diagrams})
to the isoscalar skewed quark distribution in the pion, 
$H^{I=0}(X, \xi )$, at a low normalization point,
as functions of $X$, for a value of $\xi = 1$. (Here $t = 0$). 
{\it Dashed lines:} 
Results obtained neglecting the momentum dependence of the dynamical
quark mass, {\it cf.}\  Eqs.(\ref{H_1}) and (\ref{H_2}). 
{\it Solid lines:} The corresponding contributions obtained 
when including the form factors, $F(\partial^2 )$. Note that 
the contribution from diagram (a) is non-zero only in the 
region $-\xi /2 < X < \xi /2$. The momentum dependence of the
dynamical quark mass forces this contribution to vanish
at the end points, $X = \pm\xi / 2$.}
\label{fig_H12}
\end{figure}

The corresponding skewed distributions may be computed either
using the results for the matrix elements, Eqs.(\ref{M_1_unreg}) 
and (\ref{M_2_unreg}), and the definition Eq.(\ref{skewed_def}), 
or directly by computing the R.H.S.\ of Eq.(\ref{H_explicit}); 
both ways lead to identical results. For the isoscalar part we
find
\be
H^{I=0} (X, \xi )^{\rm (a)} &=& - \frac{1}{2} \,\, \theta 
\left( -\frac{\xi}{2} < X < \frac{\xi}{2} \right) 
\frac{2 X}{\xi} ,
\label{H_1}
\\[2ex]
H^{I=0} (X, \xi )^{\rm (b)} &=& \frac{1}{2} \left[
-\theta \left( -1 < X < - \frac{\xi}{2} \right)
\; + \; \theta \left( -\frac{\xi}{2} < X < \frac{\xi}{2} \right)
\frac{2 X}{\xi} \right.
\nonumber \\
&& \left.
\;\;\;\; + \; \theta \left( \frac{\xi}{2} < X < 1 \right) \right] ;
\label{H_2}
\ee
the total result is
\be
H^{I=0} (X, \xi) &=& \frac{1}{2} \left[
-\theta \left( -1 < X < - \frac{\xi}{2} \right) \; + \; 
\theta \left( \frac{\xi}{2} < X < 1 \right) \right] .
\label{H_tot}
\ee
The functions are shown in Fig.\ref{fig_H12}. 
One sees that the contribution from diagram (a), Eq.(\ref{H_1}),
is non-zero only in the region $-\xi/2 < X < \xi /2$.
It is absent in the forward limit, $r \rightarrow 0$. 
Note that this contribution to the skewed
distribution is discontinuous in $X$ at $\pm \xi /2$; this behavior
will be modified when taking into account the momentum dependence of
the dynamical quark mass, see below. The contribution from diagram
(b) is continuous at $X = \pm \xi /2$; this part reduces in the forward
limit, $\xi \rightarrow 0$, to the singlet quark distribution in the 
pion, Eq.(\ref{distribution_def}), which in this approximation 
(keeping only the logarithmic divergence, neglecting the form factors) 
would simply be given by
\be
q_{\rm sing }(x) &=& \theta (0 < X < 1) .
\label{q_sing_model}
\ee
\par
The result Eq.(\ref{H_tot}) is consistent with the generalized momentum 
sum rule, Eq.(\ref{em_sum_rule}). In our approach based on the 
instanton vacuum the gluon distribution is parametrically small,
$\propto (M\bar\rho )^2 \sim (\bar\rho / \bar R )^4$,
so the skewed quark distribution, Eq.(\ref{H_tot}), should saturate
the sum rule at the low normalization point. Integrating Eq.(\ref{H_tot})
we observe that, indeed,
\be
\int_{-1}^1 dX \, X \, H^{I=0} (X, \xi ) &=& 
\frac{1}{2} \left( 1 \; - \; \frac{\xi^2}{4} \right) .
\label{em_sum_rule_model}
\ee
\par
Eqs.(\ref{H_1}), (\ref{H_2}) and (\ref{H_tot}) represent the result
for the skewed distribution obtained without taking into account
the momentum dependence of the dynamical quark mass. The
discontinuity at $X = \pm \xi /2$ in the contribution (a)
to $H^{I=0}$ obtained in this approximation would violate
the factorization of the DVCS amplitude, since the hard scattering
kernel contains poles at $X = \pm \xi / 2$. However, as was shown in
Refs.\cite{PPPBGW97}, the momentum dependence of the dynamical quark
mass can not be neglected for values of $X$ near $\pm \xi / 2$, 
since in this case the integral over transverse momenta is cut
by the form factors already at momenta of order 
$M \ll \bar\rho^{-1}$. The same mechanism makes the pion 
distribution amplitude vanish at the end points \cite{PP97}.
In Fig.\ref{fig_H12} we show the two contributions (a) and (b)
to $H^{I=0}$ obtained when taking into account the form factors
$F(\partial^2 )$ (we use the simple analytic approximation of
Eq.(24) of Ref.\cite{PP97}). As expected, contribution (a)
now vanishes at $X = \pm \xi / 2$, while the
modification of contribution (b), which was continuous already 
without form factors, is only quantitative.
\par
Finally, the result for the isovector skewed distribution at 
$t\rightarrow 0$ is
\be
H^{I=1} (X, \xi) &=& H^{I=1} (X, \xi)^{\rm (b)} \;\; = \;\; 
\theta \left( -1 < X < 1 \right) ,
\label{H_isovector}
\ee
which is an even function of $X$, in agreement with $C$--invariance.
[In the forward limit this corresponds to a valence quark distribution
in the pion $q_{\rm v} (X) = \theta \left( 0< X < 1 \right)$,
{\it cf.}\ Eq.(\ref{distribution_def}), so comparing with 
Eq.(\ref{q_sing_model}) we see that in this approximation 
(no form factors) the ``sea'' quark distribution in the pion
is zero.] As in the case of contribution (b) to the isoscalar 
distribution, inclusion of the form factors $F(\partial^2 )$ does not 
change the result for the isovector skewed distribution in an essential way, 
except for forcing the distribution to vanish at $X = \pm 1$.\footnote{The 
valence quark distribution in the pion has also been studied in the 
instanton vacuum in a somewhat different approach by Dorokhov and 
Tomio \cite{DorokhovTomio98}.}
\par
Some comments are in order concerning the calculation of
skewed and double distributions in the nucleon. In the large--$N_c$ 
limit the nucleon in the effective low--energy theory is
characterized by a classical pion field (``soliton'')
\cite{DPP88}. Quantization of 
the translational and rotational zero modes in the framework of the 
$1/N_c$--expansion gives rise to nucleon states with definite momentum 
and spin/isospin quantum numbers. When applying this approach to
the computation of non-forward matrix elements of the type 
Eq.(\ref{me_def}), the standard $1/N_c$--expansion implies that
different components of the average momentum, $p$, and momentum transfer, 
$r$, are of different order in $N_c$ [the nucleon mass is $O(N_c )$, 
while the momentum transfer in the Breit frame is $O(N_c^0 )$].
While it is possible to compute the skewed distribution, 
Eq.(\ref{H_explicit}), in the parametric range
$X, \xi \sim 1/N_c$ \cite{PPPBGW97}, it is not possible to uniformly
obtain the matrix element Eq.(\ref{me_def}) in the whole kinematical 
range necessary to restore the double distribution. In contrast,
in the case of the pion all components of $p$ and $r$ are $O(N_c^0 )$,
making it possible to treat $p$ and $r$ on the same footing.
\section{Crossing and the two--pion distribution amplitude}
\label{sec_crossing}
\subsection{Two--pion distribution amplitude}
An interesting feature of the pion is the fact that the quantity
related to the skewed parton distribution by crossing, namely
the two--pion distribution amplitude (DA), can be measured 
in two--pion production at low invariant masses. These $2\pi$DA's
were introduced recently in the context of the
QCD description of the process $\gamma^\ast \gamma \to 2 \pi$ 
\cite{Ter}. We now establish explicitly their relation 
to the skewed quark distributions in the pion. This will allow
us to express the $t$--dependence of the lowest moments
of the skewed distribution in terms of form factors in the 
timelike region.
\par
The two--pion DA's are defined, in analogy to the skewed parton
distribution, as the matrix elements of the twist--2 operators
between the vacuum and a two--pion state:
\be
\lefteqn{
{}_{\rm out}\langle \pi^a (p_1 ) \pi^b (p_2 ) | \; \bar\psi (-z/2 ) 
\left\{ \begin{array}{c} 1 \\[2ex] \tau^c \end{array} \right\}
\hat{z} \, [-z/2 , z/2] \, \psi (z/2)
\; | 0 \rangle } && 
\nonumber \\
&=& P\cdot z \; \int_{0}^1 du \;
e^{i (u - 1/2) P\cdot z} \; 
\left\{ \begin{array}{r}
2 \delta^{ab} \; \Phi^{I=0} (u, \zeta, W^2 )
\\[2ex]
2 i \varepsilon^{abc} \; \Phi^{I=1} (u, \zeta, W^2 ) .
\end{array}
\right. \; 
\label{definition1}
\ee
The outgoing pions have momenta $p_1, p_2$, and $P \equiv p_1 + p_2$ is 
the total momentum of the final state. The generalized DA's,
Eq.~(\ref{definition1}), depend
on the following kinematical variables: the quark momentum fraction with
respect to the total momentum of the two--pion state, $z$; the variable
$\zeta \equiv (z\cdot p_1 )/(z \cdot P)$ characterizing the distribution 
of longitudinal momentum between the two pions, and the invariant 
mass of the two--pion system, $W^2 = P^2$. Also, an explicit 
representation of the DA can be written in analogy to 
Eq.(\ref{H_explicit}) [the isospin decomposition is analogous to 
Eq.(\ref{definition1})]
\be
\lefteqn{ \Phi (u, \zeta, W^2 ) \;\; = \;\;
\frac{1}{2} \int\frac{d\tau}{2\pi}
e^{-i\tau (u - 1/2) \, p \cdot n} } 
&& \nonumber \\[1ex] 
&&\times {}_{\rm out}\langle \pi (p_1 ) \pi (p_2 ) | 
\; \bar\psi (-\tau n/2 ) \,\,
\hat{n} \,\, [-\tau n/2 , \tau n/2] \, \psi (\tau n/2)
\; | 0\rangle ,
\label{Phi_explicit}
\ee
where $n \propto z$ is a dimensionless light--like vector.
\par
From $C$--parity one derives the following symmetry
properties (we do not write the argument $W^2$):
\be
\Phi^{I=0} (u, \zeta ) &=& -\Phi^{I=0}(1 - u, \zeta)
\;\; = \;\; \;\; \Phi^{I=0}(u, 1 - \zeta ), 
\nonumber \\
\Phi^{I=1} (u, \zeta) &=& \;\; \Phi^{I=1}(1 - u, \zeta)
\;\; = \;\; -\Phi^{I=1}(u, 1-\zeta ) .
\label{reflection}
\ee
The first moment of the isovector ($I = 1$) two--pion DA
is the pion e.m.\ form factor in the time-like region,
\be
\int_0^1 du \; \Phi^{I=1}(u, \zeta, W^2 )
&=& (2\zeta - 1) \, F_\pi^{{\rm e.m.}}(W^2 ) \; ,
\label{norm}
\ee
and thus scale--independent ($F_{\pi}^{{\rm e.m.}}(0) = 1$).
For the isoscalar ($I = 0$) part, however, we have the
normalization condition \cite{MVP98,Dubrovnik}
\be
\int_0^1 du \, (2 u - 1)\, \Phi^{I=0} (u, \zeta, W^2 )
&=&
-2 M_2^{(\pi )}\; \zeta ( 1 - \zeta ) \; 
F_\pi^{{\rm EMT}} (W^2) ,
\label{norm1}
\ee
where $M_2^{(\pi )}$ is the momentum fraction carried by quarks in
the pion at the given scale, and $F_\pi^{{\rm EMT}}(W^2)$ is the
form factor of the quark part of the energy momentum tensor, normalized
to $F_\pi^{{\rm EMT}}(0) = 1$. In Ref.~\cite{MVP98} this form factor 
was estimated in the instanton model of the QCD vacuum at low 
two-pion invariant mass:
\be
F_\pi^{{\rm EMT}}(W^2) &=& 1\; + \; \frac{N_c W^2}{48 \pi^2 F_\pi^2}
\; + \; \ldots \; .
\nonumber
\ee
\par
It is useful to expand the two--pion DA simultaneously
in eigenfunctions of the ERBL evolution equation \cite{ERBL} 
[Gegenbauer polynomials $C_n^{3/2}(2 u -1)$] and 
in partial waves of the produced two--pion system
[Legendre polynomials, or Gegenbauer polynomials 
$C_l^{1/2}(2 \zeta - 1)$]. Generically this decomposition
is of the form \cite{MVP98}:
\be
\Phi (u, \zeta, W^2 ) &=&
6 u (1 - u) \; \sum_{n=0}^{\infty}
\sum_{l=0}^{n+1} B_{nl}(W^2 ) \; C_n^{3/2}(2 u - 1) \;
C_l^{1/2}(2 \zeta - 1) ,
\label{razhl} 
\ee
where $n$ runs over even (odd) and $l$ over
odd (even) integers for the isovector (isoscalar) DA,
{\it cf.}\ Eq.~(\ref{reflection}). The normalization condition 
Eq.~(\ref{norm}) requires that 
$B_{01}^{I=1}(W^2) \; =\; F_\pi^{{\rm e.m.}}(W^2)$. Note that the 
asymptotic form of the isovector two--pion DA is given by \cite{PW98}
\be
\Phi^{I=1}_{\rm asymp}(u, \zeta, W^2) &=&
6 u (1 - u) \, (2\zeta - 1) \; F_\pi^{{\rm e.m.}}(W^2) .
\label{Phi_isovector_asymp}
\ee
\par
In Ref.~\cite{MVP98} certain soft--pion theorems for the two-pion DA
were proven, which apply in the regions $\zeta \rightarrow 0$
or $\zeta \rightarrow 1$ and $W^2 \rightarrow 0$, where one of the
produced pions becomes soft. In the isovector case ($I = 1$) they
relate the two--pion DA to the DA of one pion, $\phi_\pi (u)$:
\be
\Phi^{I=1} (u, \zeta = 1, W^2 = 0 ) &=&
-\Phi^{I=1}(u, \zeta = 0, W^2 = 0 ) \;\; =\;\; \phi_\pi (u) ,
\label{letwf}
\ee
while in the the isoscalar case ($I = 0$) one obtains
\be 
\Phi^{I=0}(u, \zeta = 1, W^2 =0 ) &=&
\Phi^{I=0}(u, \zeta = 0, W^2 =0 ) \;\; =\;\; 0 .
\label{letwf0}
\ee
The theorem Eq.~(\ref{letwf}) allows to relate the expansion
coefficients of the isovector two--pion DA, Eq.~(\ref{razhl}),
at $W^2 = 0$, with those of the pion DA,
\be 
\phi_\pi (u) &=& 6 u(1-u) \; \left[ 1 \; + \; \sum_{n \; {\rm even}}
a_n^{(\pi)} \; C_n^{3/2}(2 u - 1) \right] ;
\label{razhlpi}
\ee
the relation takes the form
\be
a_n^{(\pi)} &=& \sum_{l=1}^{n+1} \; B_{nl}^{I=1}(W^2 = 0) .
\label{letcoef}
\ee
\subsection{Crossing relation}
By crossing, the matrix element defining the two--pion DA, 
Eq.(\ref{Phi_explicit}), is related to the one appearing in the definition 
of the skewed distribution, Eq.(\ref{H_explicit}). This allows one to 
express the moments of the skewed parton distribution in terms of the 
expansion coefficients of the two--pion DA, Eq.(\ref{razhl}). 
This relation takes the form:
\be
\int_{-1}^1 dX \, X^{N-1} \; H^I (X, \xi ; t)
&=& \sum_{n=0}^{N-1} \;
\sum_{l=0}^{n+1} B^I_{nl}(t) \; \left(\frac{\xi}{2}\right)^N
C_l^{1/2}\left(\frac{2}{\xi}\right)
\nonumber \\
&& \times \int_0^1 du \; 
6 u (1 - u) \,  (2 u - 1)^{N-1} \, C_n^{3/2} (2 u - 1) .
\label{relation}
\ee
One immediately notes that the R.H.S.\ is a polynomial of degree
(at most) $N$ in $\xi$, {\it i.e.}, the polynomiality condition
for the skewed distribution (see Subsection \ref{subsec_pion} and
Ref.\cite{JiReview}) is satisfied. Since only the $B_{nl}$ with
$n$ odd (even) are non-zero in the isoscalar (isovector) case,
the skewed distribution is an odd (even) function in $X$, in
agreement with $G$--parity, Eq.(\ref{symmetry}). Also, note
that due to the restrictions in the values of $l$, {\it cf.}\ 
Eq.(\ref{razhl}), $H$ is an even function of $\xi$ for both
$I = 0$ and $1$, as it should be.
\par
To prove the relation Eq.(\ref{relation}), we consider the
expression for the $N-$th moment of the skewed distribution
as a non-forward matrix element of a local spin--$N$, twist--2
operator,
\be
\lefteqn{\int_{-1}^1 dX\, X^{N-1} \;  H(X,\xi,t) }
&& \nonumber \\
&=& (2 p\!\cdot\! n )^{-N} \,
\langle \pi (p + r/2) | \bar \psi \, n\!\cdot\!\gamma 
\; (n\cdot\nablaboth )^{N-1} 
\, \psi | \pi (p - r/2) \rangle .
\label{cross1}
\ee
The $N$--th moments of the two--pion DA, Eq.(\ref{Phi_explicit}), is 
given by the vacuum to two--pion matrix element of the same local 
operator. Substituting the double expansion, Eq.(\ref{razhl}), we 
obtain
\be
\lefteqn{
\int_0^1 du \, (2 u - 1)^{N-1} \; \Phi(u, \zeta, W^2) }
&& \nonumber \\[1ex]
&=& \left[(p_1 + p_2 ) \cdot n \right]^{-N} \;
\langle \pi (p_1 ) \pi (p_2 ) | \bar\psi \, n\!\cdot\!\gamma \, 
(n\cdot\nablaboth )^{N-1} \, \psi | 0 \rangle 
\nonumber \\[1ex]
&=& \sum_{n=0}^{N-1} \; 
\sum_{l=0}^{n+1} \; B^I_{nl} \left( (p_1 + p_2 )^2 \right) \;
C_l^{1/2}\left(\frac{n\!\cdot\! (p_2 - p_1 )}{n\!\cdot\! (p_2 + p_1)} 
\right)
\nonumber \\
&& \times \int_0^1 du \; 
6 u (1 - u) \,  (2 u - 1)^{N-1} \, C_n^{3/2} (2 u - 1) .
\label{cross2}
\ee
The matrix elements of the local operators are related to 
each other by usual crossing symmetry,
\be
\langle p'| \bar \psi \, n\!\cdot\!\gamma \; (n\cdot\nablaboth )^{N-1} 
\, \psi |p\rangle 
&=& 
\langle p, -p' | \bar\psi \, n\!\cdot\!\gamma \, 
(n\cdot\nablaboth )^{N-1} \, \psi | 0 \rangle .
\ee
[In this shorthand expression, both sides are regarded as functions of 
the pion four--momenta, defined in the respective physical regions,
and analytic continuation is implied.] Using this relation with
Eqs.(\ref{cross1}) and (\ref{cross2}) we obtain Eq.(\ref{relation}). 
Note that on the R.H.S.\ of 
Eq.(\ref{relation}) the coefficients $B_{nl}(t)$ are taken at 
negative argument ($t < 0$), whereas in the expansion of the
two--pion DA, Eq.(\ref{razhl}), they are defined for positive $W^2$.
The corresponding analytic continuation can be accomplished with 
help of dispersion relations (see Ref.\cite{MVP98} for details),
\be
B_{nl}^I (t) &=& \sum_{k=0}^{N-1} \frac{t^{k}}{k!}\frac{d^k}{dt^{k}}
B_{nl}^I (0) \; + \;
\frac{t^{N}}{\pi}\int_{4m_\pi^2}^\infty
ds \frac{\tan\delta_l^I (s)\,\mbox{Re}\, B_{nl}^I(s)}{s^N(s-t-i0)} ,
\label{dr}
\ee
where $\delta_l^I (s), \; (I = 0, 1)$ are the $\pi\pi$ scattering phase 
shifts in the isospin 0 and 1 channels.
\par
Let us see the implications of the crossing relation, 
Eq.(\ref{relation}), for the lowest moments of the skewed 
distribution. For the first moment of the isovector distribution
we have
\be
\int_{-1}^1 dX  \, H^{I=1}(X,\xi; t) &=& B_{01}(t) \;\; = \;\;
F_\pi^{{\rm e.m.}}(t),
\ee
where $F_\pi^{{\rm e.m.}}(t)$ is the pion electromagnetic form factor
in the spacelike region, in agreement with Eq.(\ref{norm_isovector}). 
For the second moment of the isoscalar 
distribution we obtain, substituting the explicit form of the
Gegenbauer polynomials in $\xi/2$:
\be
\int_{-1}^1 dX \, X \; H^{I=0}(X,\xi,t) &=&
\frac{3}{5} \left[ B_{10}(t) \frac{\xi^2}{4} \; + \; 
B_{12}(t)\frac{12-\xi^2}{8}
\right] .
\ee
At $t=0$, using the soft--pion theorem, Eq.(\ref{letwf0}), 
which implies $B_{10}(0) \, + \, B_{12}(0) \; = \; 0$, we get
\be
\int_{-1}^1 dX \, X \; H^{I=0}(X,\xi; t=0)
&=& \frac{9}{10} B_{12}(0) \left( 1 - \frac{\xi^2}{4} \right) .
\ee
If we substitute the value $B_{12}(0) = 5/9$, which was 
computed in Ref.\cite{MVP98} within the instanton vacuum model,
we obtain precisely the generalized momentum sum rule, 
Eq.(\ref{em_sum_rule}).
\subsection{Application to quark/antiquark distributions in pion}
An particularly interesting application of the crossing relation, 
Eq.(\ref{relation}), is the forward limit, $t = 0$ and $\xi = 0$, 
where the skewed distributions reduce to the usual quark/antiquark
distributions, {\it cf.}\ Eq.(\ref{distribution_def}).
The quark/antiquark distribution are known with fair accuracy
from parametrizations of $\pi N$ Drell--Yan and other 
data \cite{GRV92}. Eq.(\ref{relation}) relates the moments
of the quark/antiquark distributions in the pion to the expansion 
coefficients $B_{nl}$ at $t = 0$, which can in principle be measured
in two--pion production at low invariant masses \cite{MVP98}.
The relation takes the form [{\it cf.}\ Eq.(\ref{distribution_def})]
\be
M_N^{(\pi)} &\equiv& \int_{0}^1 dX\, X^{N-1} 
\left\{ \begin{array}{r} q_{\rm v} (X) \\[2ex]
2 q_{\rm s} (X)
\end{array} \right\} 
\;\; = \;\; A_N 
\left\{ \begin{array}{lr} 
B^{I=1}_{N-1,N}(0) & \mbox{$N$ odd} ,\\[2ex]
B^{I=0}_{N-1,N}(0) & \mbox{$N$ even} ,
\end{array} \right.
\label{f2tp}
\ee
where the $A_N$ are numerical coefficients which can be determined
from Eq.(\ref{relation}): $A_1 = 1, \; A_2 = 9/5, \;
A_3 = 6/7, \; A_4 = 5/3$, {\it etc}.
For the lowest moments Eq.(\ref{f2tp}) implies 
$B^{I=1}_{01}(0) \, = \, M_1^{(\pi)} \, = \, 1$, which corresponds to 
the normalization condition Eq.(\ref{norm}), and 
$B^{I=0}_{12}(0) \, = \, 5/9 \, M_2^{(\pi)}$, which corresponds 
to Eq.(\ref{norm1}).\footnote{To see this one needs to use the 
soft--pion theorem for the isoscalar two--pion DA, 
Eq.(\ref{letwf0}).} 
\par
A non-trivial relation is obtained for $N = 3$. Using Eq.(\ref{f2tp}) 
and the soft--pion theorem for the isovector two--pion DA, 
Eq.(\ref{letwf}), we can determine the coefficient $B_{21}(0)$
describing the deviation of the isovector two--pion DA from 
its asymptotic form, {\it cf.}\ Eq.(\ref{Phi_isovector_asymp}).
One finds
\be
B_{21}^{I=1}(0) &=& a_2^{(\pi)} \; - \; \frac{7}{6} M_3^{(\pi)} ,
\label{B_21_from_M_3}
\ee
where $a_2^{(\pi)}$ is the expansion coefficient for the
pion DA, Eq.(\ref{razhlpi}). This relation is model independent
an can be used as a consistency check for model calculations.
\par
Computation of the valence quark distribution in the pion in the 
effective theory based on the instanton vacuum 
(see Section \ref{sec_model}) gives a value of 
$M_3^{(\pi)} = 0.25$. [In this calculation the instanton--induced
form factors $F(\partial^2 )$ have been taken into account.] 
The second moment of the pion DA has been
computed in the same approach in Ref.\cite{PPRGW98}, 
$a_2^{(\pi)} = 0.062$; this small value is consistent with the
CLEO measurements \cite{CLEO98}. Substituting these results
in Eq.(\ref{B_21_from_M_3}) we obtain $B_{21}^{I=1}(0) = -0.22$, 
which agrees with the result of a direct calculation in the
instanton vacuum in Ref.\cite{MVP98}.\footnote{The slight numerical
difference with the result quoted in Ref.\cite{MVP98}, 
$B_{21}^{I=1}(0) \approx -0.20$, is due to different approximations
used in treating the form factors $F(\partial^2 )$.} Thus, we see that the
results obtained from the effective theory based on the instanton
vacuum are consistent with the relation Eq.(\ref{B_21_from_M_3}).
This is an extremely non-trivial check, since it shows that this
approach preserves the soft--pion theorems ({\it i.e.}, 
chiral invariance) as well as crossing symmetry.
\par
We note that the value of $M_3^{(\pi)} = 0.25$ obtained from the
instanton vacuum is somewhat larger than that of the GRV 
parametrization at the low normalization point \cite{GRV92}, 
$M_3^{(\pi)} = 0.16$, and in good agreement with the value 
extracted from QCD sum rules with non-local condensates by Belitsky, 
$M_3^{(\pi)} \approx 0.29$ \cite{Belitsky96}.
\par
When the $\pi\pi$ phase shifts in Eq.(\ref{dr}) in the isovector channel 
are approximated by exchange of ``elementary'' resonances 
($\rho , \rho' , \ldots$), it becomes possible to express 
the expansion coefficients of the isovector two--pion DA in terms 
of the moments of the distribution amplitudes of the resonances
(see Ref.\cite{MVP98} for details). Keeping only the dominant
contribution from $\rho$ exchange in Ref.\cite{MVP98} was 
obtained the relation
\be
a_2^{(\rho)} &=& B_{21}^{I=1}(0) \, e^{-C M_\rho^2} ,
\ee
where the coefficient $C$ was estimated in the instanton vacuum,
$C \approx 0.6 \, {\rm GeV}^{-2}$. In this approximation
Eq.(\ref{B_21_from_M_3}) becomes
\be
a_2^{(\rho)}  e^{C M_\rho^2} &=& a_2^{(\pi)} 
\; - \; \frac{7}{6} M_3^{(\pi)} .
\label{a2_from_M_3}
\ee
Apart from the value of $C$, which does not influence the sign of
the L.H.S., this relation is again model independent, and we can
use it as a test for models of resonance DA's. We already
noted that the instanton vacuum predicts
$B_{21}^{I=1}(0) \approx - 0.2$, and thus a negative value for
$a_2^{(\rho )}$, which is consistent with Eq.(\ref{a2_from_M_3})
because of the small value for $a_2^{(\pi)}$ obtained in
this approach; see above. This result for $a_2^{(\rho )}$
is in contradiction to the results of QCD sum rule 
calculations, both in the standard approach \cite{BB96} and 
with non-local condensates \cite{BakMikh98}, which obtained
positive values. Ref.\cite{BB96} reported a value of 
$a_2^{(\rho)} = 0.18 \pm 0.1$. However, this calculation
appears to be consistent with Eq.(\ref{a2_from_M_3}), since
a comparable sum rule calculation of $a_2^{(\pi)}$ \cite{BF89} 
arrives at a relatively large value of $a_2^{(\pi)} \approx 0.44$, 
so that Eq.(\ref{a2_from_M_3}) is satisfied if one  
substitutes, say, the value for $M_3^{(\pi)}$ from the GRV 
parametrization \cite{GRV92}, or a slightly larger one. 
On the other hand, QCD sum calculations
with non-local condensates give a significantly smaller value
for the second moment of the pion DA; Ref.\cite{MikhRad92}
estimates $a_2^{(\pi)} = 0 \ldots 0.15$; see also
Ref.\cite{BakMikh95}. $M_3^{(\pi)}$ was estimated in 
a QCD sum rule calculation with non-local condensates in 
Ref.\cite{Belitsky96}, $M_3^{(\pi)} \approx 0.29$. This could
indicate that in the calculation of $a_2^{(\rho)}$ in 
Ref.\cite{BakMikh98}, which quotes a value of $0.08 \pm 0.02$,
the error margin could be somewhat larger than estimated. 
\section{Conclusions}
\label{sec_conclusions}
In this paper we have investigated the structure of non-forward
matrix elements of light--ray operators at a low normalization point,
and their representations in terms of skewed and double distributions.
Our principal conclusion is that the skewed distribution generally
has a ``two--component'' structure, {\it i.e.}, that one is 
dealing with essentially different functions in the 
``quark/antiquark distribution'' region, 
$X > \xi / 2, X < -\xi / 2$,
and the ``meson distribution amplitude'' region,
$-\xi/2 < X < \xi / 2$. This follows naturally 
from the double distribution representation of the matrix element
if one explicitly includes {\it all} twist--2 structures in the 
double distribution representation. We have shown that the 
contributions resulting from the ``longitudinal'' ($\propto r_\mu$) 
twist--2 structure cannot be obtained from delta function terms
in the conventional double distribution, which have previously been
discussed by Radyushkin \cite{Rad2,Rad3}.
\par
This conclusion concerning the behavior of the skewed distribution
at a low normalization point does not depend on any assumptions about 
the details of the non-perturbative at low scales. In fact, we have found
qualitatively similar behavior in two different dynamical models: 
{\it i)} the low--energy effective theory based on the instanton 
vacuum, and {\it ii)} a generic meson exchange model. 
Moreover, the necessity to include terms 
$\propto r_\mu$ in the decomposition of the non-forward matrix element 
Eq.(\ref{vector_operator}) is revealed by crossing invariance, which 
relates these terms to the matrix element for production of two pions.
\par
Our results concerning the general structure of non-forward matrix 
elements can serve as a basis for the construction of realistic 
parametrizations of skewed distributions, satisfying all known
requirements, and reproducing the phenomenologically known
quark/antiquark and gluon distributions as well as the form 
factors of local operators in the appropriate limits. We 
plan to address this topic in a future publication. Whether
or not the double distribution representation, in its complete
form, Eq.(\ref{double_minimal}), will prove to be a useful tool
for modeling skewed distributions remains to seen.
\par
The crossing relations between the skewed quark distribution
and the two--pion distribution amplitude derived here can be 
used to obtain additional information about the quark/antiquark 
distribution in the pion from measurements of electroproduction of 
two pions in $\gamma^\ast \gamma$ and $\gamma^\ast N$ reactions. This
fundamental characteristic of the pion is up to know only poorly 
known, since it can be measured directly only in hadronic $\pi N$
reactions such as Drell--Yan production, where it enters always 
together with the (anti--) quark distributions in the nucleon. 
In particular, two--pion production in $\gamma^\ast \gamma$ reactions 
provides an opportunity to measure the distributions in the pion in 
a purely electromagnetic process.
\\[1cm]
{\large\bf Acknowledgements} \\[.2cm]
We are grateful to L.\ Mankiewicz, A.V.\ Radyushkin, A.\ Sch\"afer,
and A.G.\ Shuvaev for discussions of properties of skewed 
and double distributions, and to V.M.\ Braun, P.\ Ball, and A.V.\ Belitsky 
for other valuable communication. We also acknowledge many interesting
discussions with A.P.\ Bakulev, K.\ Goeke, V.Yu.\ Petrov, 
P.V.\ Pobylitsa, R. Ruskov, N. Stefanis, and O.\ Teryaev.
\\[.2cm]
This work has been supported in part by a
joint grant of the Russian Foundation for Basic Research (RFBR) and the
Deutsche Forschungsgemeinschaft (DFG) 436 RUS 113/181/0 (R), by
RFBR grant 96-15-96764, by the NATO Scientific Exchange grant
OIUR.LG 951035, by the DFG and by COSY (J\"ulich).
\newpage
\appendix
\renewcommand{\theequation}{\Alph{section}.\arabic{equation}}
\section{Generalized momentum sum rule for the pion}
\label{app_em}
\setcounter{equation}{0}
In this appendix we give a derivation of the generalized momentum sum 
rule for the isoscalar skewed distribution in the pion, 
Eq.(\ref{em_sum_rule}), using the universal chiral Lagrangian.
\par
The sum of the second moments of the skewed quark and gluon 
distributions is related to the form factors of the energy--momentum 
tensor \cite{Ji2}. The symmetric version of the QCD EM--tensor
is given by
\be
T_{\mu\nu} (0) &=& \frac{1}{2} \bar\psi (0)
\gamma_{\left\{ \mu \right.} \nablaboth_{\left. \nu \right\}}
\psi (0) 
\; + \; F_{\mu\alpha} (0) F^\alpha_\nu (0) 
\; - \; \frac{1}{4} \, g_{\mu\nu} \, F^2 .
\ee
The general form of its matrix element between pion states is
\be
\langle \pi^+ (p - r/2 ) | \; T_{\mu\nu} (0) \;
\; | \pi^+ (p + r/2 ) \rangle &=& A(t) p_\mu p_\nu
+ B(t) (r_\mu r_\nu - r^2 g_{\mu\nu})
\label{em_general}
\ee
($t = r^2$); all other structures vanish due to energy--momentum
conservation, $\partial_\mu T_{\mu\nu} (x) = 0$. Expanding the
non-local operators in Eqs.(\ref{me_def}) and (\ref{me_gluon}) 
in the distance $z$,
\be
\bar\psi (-z/2 ) \hat{z} \, [-z/2, z/2] \, \psi (z/2) &=&
\bar\psi (0) \hat{z} \psi (0) + \frac{1}{2} z^\mu z^\nu
\bar\psi (0) \gamma_\mu \nablaboth_\nu \psi (0) + \ldots \;\;\;\;\;
\\
z^\mu z^\nu F_{\mu\alpha} (-z/2 ) \, [-z/2, z/2] \,
F^\alpha_\nu (z/2) &=&
z^\mu z^\nu F_{\mu\alpha} (0) F^\alpha_\nu (0) + \ldots 
\ee
and keeping in mind that $\rz = \xi \pz$, one can easily show that
\be
\int_{-1}^1 dX \, X \, [H^{I=0} + H^G] (X, \xi; t) &=& \frac{1}{2}
A(t) + \frac{\xi^2}{2} B(t) .
\label{H_A_B}
\ee
\par
At $t \rightarrow 0$ the matrix element of the energy--momentum tensor
can be computed from first principles, since soft pion dynamics is 
described by the universal chiral Lagrangian. From
\be
L_{\rm soft-pion} &=& \frac{1}{2} (\partial_\mu \pi)^2
\label{L_soft}
\ee
one obtains
\be
T_{\mu\nu, {\rm soft-pion}} &=&
\partial_\mu \pi \partial_\nu \pi - \frac{1}{2}
g_{\mu\nu} (\partial_\mu \pi)^2 .
\ee
Calculation of the matrix element between pion states,
{\it cf.}\ Eq.(\ref{em_general}), gives
\be
A_{\rm soft-pion} (t = 0) &=& 2, \hspace{2cm}
B_{\rm soft-pion} (t = 0) \;\; = \;\; -\frac{1}{2} .
\ee
Inserting this in Eq.(\ref{H_A_B}) for $t \rightarrow 0$
one obtains Eq.(\ref{em_sum_rule}).
\section{Resonance exchange contributions to non-forward pion matrix element}
\label{app_resonance}
\setcounter{equation}{0}
An interesting property of non-forward matrix elements of QCD
operators is the possibility of ``meson exchange'' contributions.
One may view the exchange of a $t$--channel resonance as a purely
phenomenological model for the matrix elements at a low scale, 
in the spirit of the vector dominance picture for the pion and nucleon 
e.m.\ form factors \cite{Sakurai60}. Also, at large--$N_c$, where 
QCD is believed to become equivalent to a theory of resonances,
one may hope to eventually construct a complete description
of the pion matrix element of QCD operators in terms of resonance
exchange.
\par
In this appendix we compute the contribution of the $t$--channel 
exchange of a spin--$J$ two--pion resonance to the non-forward
matrix element of the twist--2 operator in the pion, Eq.(\ref{me_def}).
Specifically, we want to show that exchange of isoscalar resonances
(even $J$) gives rise to the behavior of the matrix element stated
in Eq.(\ref{M_nonzero}).
\par
\begin{figure}
\setlength{\epsfxsize}{10cm}
\setlength{\epsfysize}{10cm}
\epsffile{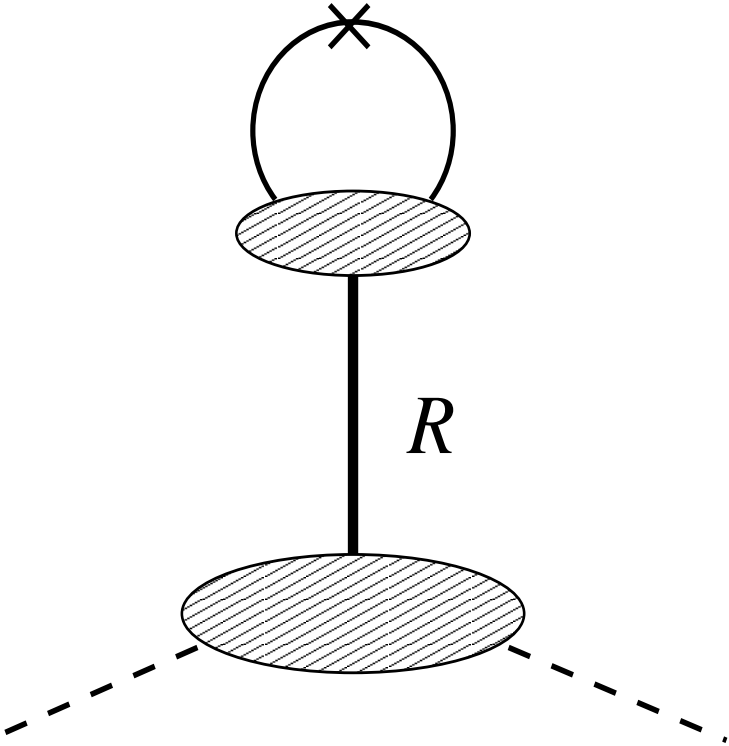}
\caption[]{Schematic representation of resonance exchange contributions
to the non-forward matrix element of the twist--2 operator in the
pion, Eq.(\ref{me_def}). The upper blob denotes the distribution
amplitude of the spin--$J$ resonance, Eq.(\ref{R_da}).}
\label{fig_exchange}
\end{figure}

The contribution to the pion matrix element of the 
light--ray operator, Eq.(\ref{vector_operator}), 
from an exchange of a resonance of spin $J$ 
is given by the amplitude for the pion to emit the resonance,
the resonance propagator, and the matrix element for the
resonance to be ``absorbed'' by the light--ray operator
(see Fig.\ref{fig_exchange}).
The coupling of the resonance to the pion is of the form
\be
\langle \pi (p - r/2 ) R (r, J, \lambda ) 
| \pi (p + r/2 ) \rangle
&=& g_{R\pi\pi} \; p^{\rho_1} \ldots p^{\rho_n} 
\epsilon_{\rho_1 \ldots \rho_n}^{(\lambda )\ast} ,
\label{R_emission}
\ee
with isospin structure analogous to that of Eq.(\ref{me_def}).
Here $\epsilon_{\nu_1 \ldots \nu_n}^{(\lambda)}$ denotes
the polarization tensor of the spin--$J$ resonance, and the coupling 
constant, $g_{R\pi\pi}$, can be related to the $\pi\pi$ width of the 
resonance. The upper part of the diagram in Fig.\ref{fig_exchange} is 
the matrix element
\be
\lefteqn{
\langle 0 |
\bar\psi (-z/2) \hat{z} \, [-z/2, z/2] \,
\psi (z/2) | R(r, J, \lambda ) \rangle }
&& \nonumber \\[1ex]
&=&  f_R \; (M_R )^J  
\frac{\epsilon_{\nu_1 \ldots \nu_J}^{(\lambda)}
z^{\nu_1} \ldots z^{\nu_J}}{(\rz )^{J - 1}} 
\; \int_{-1}^1 dy\; e^{-i y \rz /2} \; \phi_R (y) .
\label{R_da}
\ee
Here, $\phi_J (y)$ is a twist--2 distribution amplitude of the 
spin--$J$ resonance. Conservation of angular momentum implies
that 
\be
\int_{-1}^1 dy\, y^n \, \phi_R (y) &=& 0 \hspace{1cm}
\mbox{for} \;\;\;\;
\left\{\begin{array}{rr} 
0 \leq n \leq J - 1 & \mbox{$J$ odd} , \\
1 \leq n \leq J & \mbox{$J$ even} .
\end{array}
\right.
\ee
Computing the diagram Fig.\ref{fig_exchange} we obtain for the resonance
contribution to Eq.(\ref{me_def}):
\be
\lefteqn{
\cM (\pz , \rz ; t)_{R-{\rm exch}}  \;\; = } && \nonumber \\[2ex]
&& \sum_R
\frac{f_R \, g_{R\pi\pi} \, (M_R )^J}{M_R^2 - t}  \;
\frac{(-1)^J \, J!}{(2J - 1)!!}
\; \rz \; P_J \!\left( \frac{2 \pz}{\rz} \right) 
\; \int_{-1}^1 dy\; e^{-i y \rz /2} \phi_R (y) ,
\label{R_M}
\ee
where $P_J$ is the Legendre polynomial of degree $J$, which results
from the contraction of the vectors $p$ and $z$ with the
transverse projector
\be
p^{\rho_1} \ldots p^{\rho_J} z_{\nu_1} \ldots z_{\nu_J} 
\sum_\lambda \epsilon^{(\lambda)\; \nu_1 \ldots \nu_n}
\epsilon_{\rho_1 \ldots \rho_n}^{(\lambda )\ast} .
\ee
\par
In particular, in the limit $\pz \rightarrow 0$ the
argument of the Legendre polynomial in Eq.(\ref{R_M})
becomes zero.
Since $P_J(0) = 0$ for odd $J$ we see that the exchange of isovector
(odd--$J$) resonances does not contribute to the value of the
amplitude at $\pz \rightarrow 0$, in accordance with the
property Eq.(\ref{symmetry}) of $\cM^{I = 1}$ discussed in 
Section \ref{sec_general}.
In the isoscalar case, however, Eq.(\ref{R_M}) is non-zero.
This is precisely the behavior described in Eq.(\ref{M_nonzero}),
which makes a double distribution representation in the 
form Eq.(\ref{double_naive}) impossible. In fact, in the modified 
representation of $\cM^{I=0} (\pz , \rz ; t)$, Eq.(\ref{double_minimal}),
isoscalar exchange leads to a contribution described by
\be
D (y)_{R-{\rm exch}} &=& 
\sum\limits_{\scriptstyle R\atop \scriptstyle J\; {\rm even}}
\frac{f_R \, g_{R\pi\pi} \, (M_R )^J}{M_R^2 - t} \; 
\frac{(-1)^{J/2}\, [(J - 1)!!]^2}{(2J - 1)!!} \; \phi_R (y) .
\label{R_F_r}
\ee
From Eq.(\ref{R_M}) one sees that exchange of a $J = 0$ resonance 
(``sigma meson'') contributes only to the $\rz$--term in the
double distribution representation, Eq.(\ref{double_minimal}),
and thus only to $D (y)$, while even--spin resonances with 
$J \ge 2$ contribute to both $F^{I=0}(x, y)$ and $D(y)$.
%
%

%
%
\newpage
\newpage
\newpage
\newpage
\newpage
\begin{figure}
\setlength{\epsfxsize}{15cm}
\setlength{\epsfysize}{15cm}
\epsffile{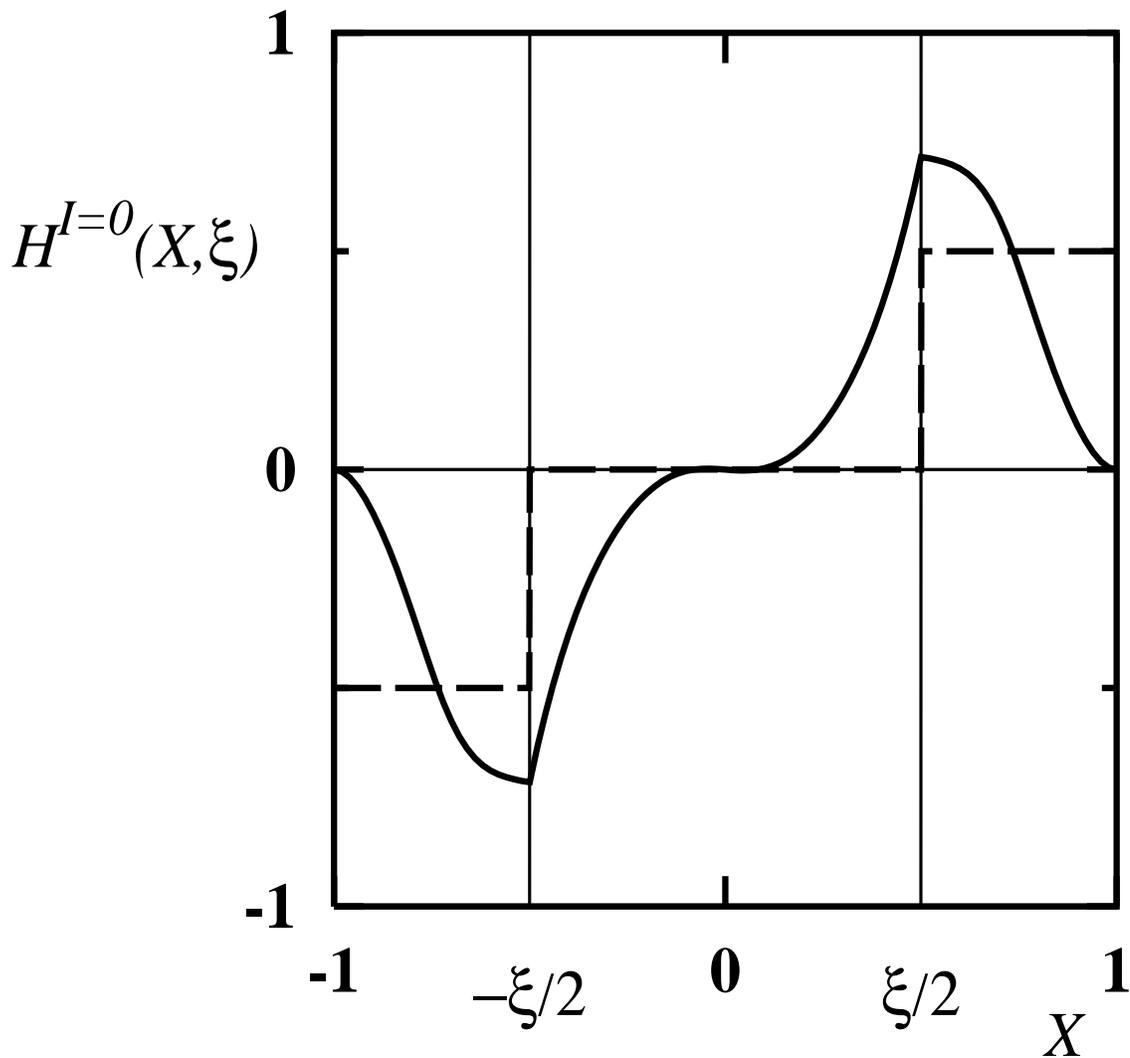}
\caption[]
{The total isoscalar skewed quark distribution in the pion,
$H^{I=0} (X, \xi )$, for $\xi = 1$, being the sum of the two 
contributions (a) and (b) shown in Fig.\ref{fig_H12}.
{\it Dashed line:} Result obtained neglecting the momentum 
dependence of the dynamical quark mass, {\it cf.}\ Eq.(\ref{H_tot}). 
{\it Solid lines:} Distribution obtained when including the 
form factors, $F(\partial^2 )$. Due to the vanishing of contribution (a)
at $X = \pm \xi/2$ the total distribution is continuous
at these points.}
\label{fig_Htot}
\end{figure}
\end{document}